\documentclass[journal=jacsat,manuscript=article]{achemso}
\usepackage [latin1]{inputenc}
\usepackage{achemso}
\setkeys{acs}{articletitle=true}
\setkeys{acs}{maxauthors = 0}
\usepackage{chemformula} % Formula subscripts using \ch{}
\usepackage[T1]{fontenc} % Use modern font encodings
\usepackage{epsfig}
\usepackage{graphicx}
\usepackage{dcolumn}
\usepackage{amsthm,amsmath}
\usepackage{pstricks}
\usepackage{color}

\author{Ravi Kumar}
\affiliation{Academy of Scientific and Innovative Research, CSIR-Human Resource Development Center (CSIR-HRDC) Campus, Postal Staff College Area
,Ghaziabad, Uttar Pradesh, 201002, India}
\alsoaffiliation[CSIR-National Chemical Laboratory]
{Electronic Structure Theory Group, Physical Chemistry Division, CSIR-National Chemical Laboratory, Pune, 411008, India}

\author{Aryya Ghosh}
\affiliation[Ashoka University]
{Department of Chemistry, Ashoka University, Sonipat, Haryana,131029 India}

\author{Nayana Vaval}
\email{np.vaval@ncl.res.in}
\affiliation[CSIR-National Chemical Laboratory]
{Electronic Structure Theory Group, Physical Chemistry Division, CSIR-National Chemical Laboratory, Pune, 411008, India}

\title[An  \textsf{achemso} demo]
  {Decay  processes in microsolvated clusters : A complex  absorbing  potential  based equation-of-motion  coupled  cluster investigations\footnote{Decays in cationic alkali metals}}

\abbreviations{CC,EOM,ICD,ETMD,Auger decay}

\begin{document}

 \begin{tocentry}
    \centering
   \includegraphics[width=0.7\textwidth]{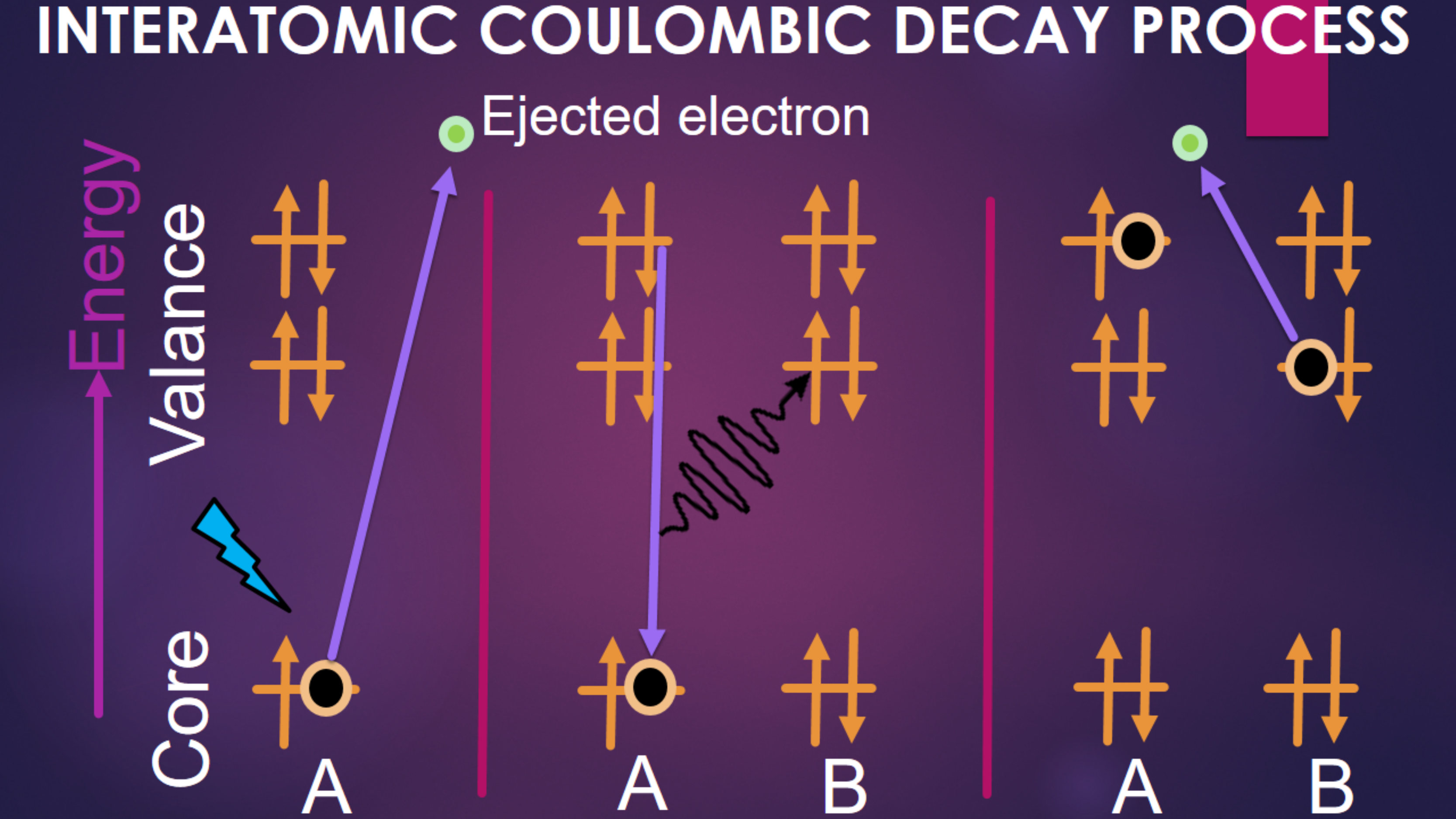}

 \end{tocentry}

\begin{abstract}
  We have  employed the highly accurate complex absorbing  potential  based  ionization potential equation-of-motion coupled  cluster singles and doubles (CAP-IP-EOM-CCSD) method to  study the various  intermolecular  decay  processes in ionized metals  (Li$^{+}$, Na$^{+}$, K$^{+}$) microsolvated  by water  molecules. For  the  Li  atom, the electron  is ionized from the 1s subshell. However, for Na and K atoms, the electron is  ionized from  2s  and both 2s and 2p subshells, respectively.
We have  investigated  decay  processes  for  the  Li$^{+}$-(H$_{2}$O)$_{n}$; (n=1-3) systems as well as Na$^{+}$-(H$_{2}$O)$_{n}$; (n=1,2), and K$^{+}$-H$_{2}$O. The Lithium cation in water can decay only via electron transfer mediated decay (ETMD) as there are no valence electrons in Lithium.  We have  investigated  how  the various decay  processes change in the  presence of  different alkali  metal atoms and how  the increasing  number  of  water molecules  play  a  significant role in the decay of microsolvated systems. To see the effect of the environment, we have studied the Li$^{+}$-NH$_{3}$ (in comparison to Li$^{+}$-H$_{2}$O). In the case of Na$^{+}$-H$_{2}$O, we have studied the impact of bond distance on the decay width. The effect of polarization on decay width is checked for the X$^{+}$-H$_{2}$O; X=Li, Na. We have used the PCM model to study the polarization effect. We have compared  our results  with  the  existing theoretical and experimental  results  wherever available  in the  literature.
\end{abstract}

\section{Introduction}

There are various ways in which an excited/ionized atom/molecule can relax. It  can  relax  either  via  radiative processes or 
non-radiative  decay  processes.  
A radiative decay like fluorescence or a non-radiative process
like Auger decay has been exceptionally well  known for a long time. Auger spectroscopy \cite{Auger} has various applications in 
material and surface science.
In 1997, Cederbaum \textit{et al.} \cite{ICD,ICD1} proposed a new decay mechanism for  inner valence ionized/excited  states, called
interatomic/molecular Coulomb decay (ICD).  This non-local complex relaxation process happens in atomic/molecular clusters. In the original
formulation of this process, a single
inner-valence hole state in an atom or molecule, which cannot decay locally via the Auger
mechanism due to energetic considerations, decays through energy transfer to the neighbouring atom. 
This knocks out an outer valence electron from the adjacent atom or molecule. Contrary to the Auger decay process, ICD is driven by the 
correlation between electrons located on different species, often a few nanometers
apart. The ejected ICD electrons have low energy, whose value strongly depends on the
initial state and chemical nature of the neighbour. In the ICD process,  two positive
charges are produced in close proximity of each other, leading to a Coulomb explosion.

Electron transfer mediated decay (ETMD) \cite{etmd} is another interatomic decay process initiated by ionizing radiation.
In this process, not energy but electron transfer between two
sub-units acts as a mediator. In ETMD, a neighbour donates an electron to an initially ionized
atom or molecule, while excess energy is transferred either to the donor or to another neighbour, which emits a secondary 
electron to the continuum. Li$^+$ has only electrons in its core orbital, so it is the best example to study the ETMD process. Being an electron transfer process, 
ETMD is usually considerably slower than energy transfer driven ICD process. However, it
becomes a vital decay pathway in a medium if ICD is energetically forbidden.
Since its original formulation, ICD has been investigated theoretically \cite{ICDtheo,icdtheo1,icdtheo2} and experimentally
\cite{ICDexpt,ICDexpt1,ICDexpt2,ICDexpt3} in  a 
variety  of systems  such  as  rare-gas \cite {icdrare,icdhene}  clusters, hydrogen bonded  clusters \cite{icdhbond} and 
water  solutions \cite{watersol,watersol1}. It has been found that not only inner-valence ionized states may undergo ICD, but any 
localized electronic excitation whose energy lies above the ionization potential of a neighbour can undergo ICD. Thus,
the ICD of ionized-excited, doubly ionized, or neutral-excited states of clusters have been observed and investigated  theoretically. 
Moreover, it turns out that this  decay process can be initiated not only by photons but also by energetic electrons  and positive
heavy ions.
Experimentally, the ETMD process has  been  observed in  rare  gas  clusters\cite{etmdexpt,etmd2,etmdraregas,etmdr1} and  alkali doped 
helium droplets 
\cite{hedroplt,heli2,hedrop,etmdli}. Recently, Unger \textit {et al.}  \cite {unger} have  investigated  the ETMD  process in LiCl  aqueous  solution.
The ETMD  process has been  studied  theoretically in  hydrogen bonded  clusters \cite{etmdhbond,etmdhbond1}. Recently, Ghosh \textit {et al.} \cite {ghosh21} have  investigated 
the  ETMD  process 
in the HeLi$_2$ cluster. Their investigation, has  shown that the multi-mode  nuclear dynamics  play a significant
role in the ETMD  process. The ETMD  process  has  also  been  investigated  theoretically  in  microsolvated clusters \cite{microsolv}. 
  
The high efficiency of interatomic decay processes (i.e. ICD, ETMD) makes it imperative to take these interatomic
decay processes into account for proper understanding of physico-chemical phenomena induced in biological systems by ionizing radiation and related to radiation damage.  
First, ICD and ETMD result in the production of
low energy electrons (LEEs) through ionization of the medium. LEEs are known to be very
effective in causing DNA strand breaks through the resonant dissociative attachment mechanism \cite{brun}. Second,
the ionization of the medium produces genotoxic radicals such as the hydroxyl
radical OH. Third, both LEE and radicals are produced locally close to where an ionizing particle initially deposits energy. If this happens close to the DNA molecule,
the probability of the complex being damaged significantly increases. The feasibility of ICD and
ETMD among biologically relevant species was investigated theoretically \cite{icdbiomol,icdbio2}.
 Microsolvated  clusters  with alkali metal cations serve as  a model  system  for a natural  biological  system because ions can impact the intracellular and extracelluar activities, i.e. the movement of enzymes \cite{enzyme} and activities and conformers of proteins \cite{proteinfolding}. Na$^+$ and K$^+$ ions are the key component of the sodium-potassium pump  in the human body. These ions also help to transmit the signals inside the brain \cite{brainsignals}. The proper functioning of Na$^+$ and K$^+$ ions helps us to avoid neurological diseases \cite{braindis}. Therefore, the  investigation  of  interatomic  or  intermolecular  decay  processes  in  microsolvated  clusters  will  shed light  on  chemistry  related to radiation  damage.
 The decay rate of the interatomic decay  process specifically depends on the energy of the initially ionized or excited state. Therefore,
the proper treatment of the initially ionized or excited state is necessary to calculate the lifetime of the interatomic  
decay process. The ionization potential equation-of-motion coupled cluster method augmented by complex absorbing potential (CAP-IP-EOM-CC) method
\cite{capeom, capeom1, capeom2, capeom3a, capeom3b, capeom3c, capeom4} provides proper treatment of ionized states or excited states with the inclusion of 
correlation effects (dynamic and non-dynamic) as well as continuum ones. Therefore, the CAP-IP-EOM-CCSD method is promising to 
describe the interatomic  decay process efficiently.

  In this paper, we  have reported the implementation of the highly  correlated  CAP-IP-EOM-CCSD method, which  is  a combination of  the CAP approach  and  equation-of-motion  coupled  cluster \cite{eomcc,eomcc1,eomcc2,eomcc3,eomcc4,eomcc5,eomcc6} approach 
to study the ETMD decay mechanism in  microsolvated Li$^{+}$-(H$_{2}$O)$_{n}$ (n=1,3) systems, ICD in Na$^{+}$-(H$_{2}$O)$_{n}$;n=1,2 and Auger decay in K$^{+}$-(H$_{2}$O).  
To see the effect of the environment on the decay of Li 1s state, we have chosen two iso-electronic systems, i.e. Li$^{+}$-NH$_{3}$  and Li$^{+}$-H$_{2}$O. 
We compare our results with the available theoretical/experimental results wherever available. 
This paper is organized as follows; in Section 2, we briefly discuss the equation-of-motion coupled cluster theory 
along with the CAP approach.  Results and discussion on them are presented in Section 3. In Section 4, we conclude our findings.

\begin{figure}
\includegraphics[width=1.0\textwidth]{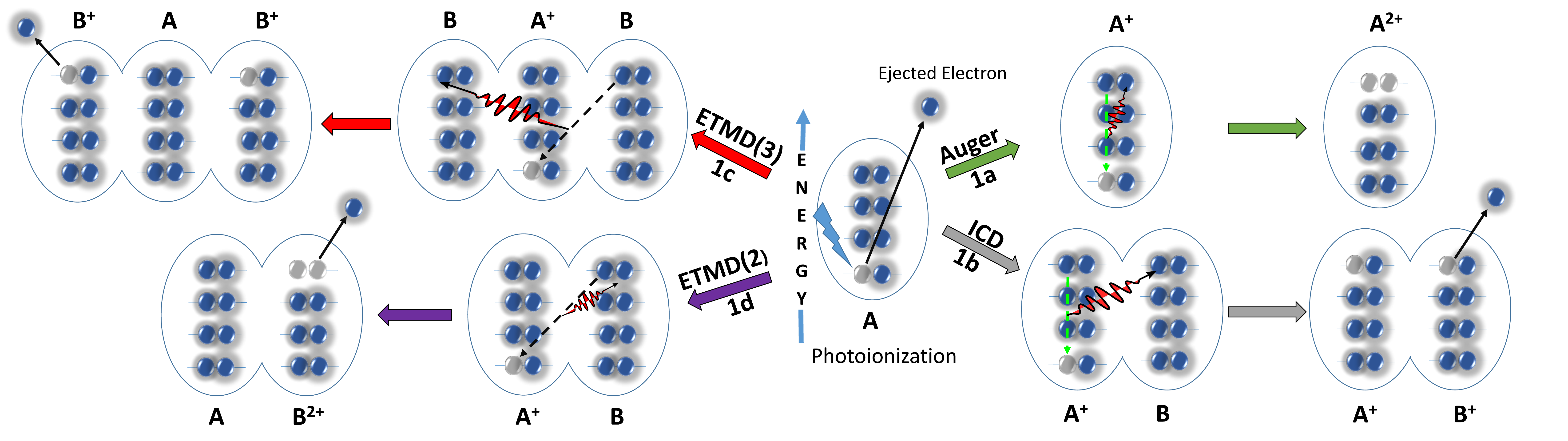}
%\vspace {-15cm}
    \caption{Various non-radiative decay processes: 1a Auger decay, 1b ICD, both 1c and 1d  ETMD processes. 3 and 2 represents the number of atoms in 1c and 1d, respectively. }
    \label{fig:effects}
\end{figure}

\section{Theory}\label{sec2}
\subsection{Complex absorbing potential based equation-of-motion coupled cluster}

To calculate the   position and lifetime of the decaying state, we have used the CAP-IP-EOM-CCSD method. In this section, we discuss the CAP-IP-EOM-CCSD method briefly. The decaying states are associated with the complex eigenvalues within the formulation of Siegert\cite{siegert}.

\begin{eqnarray}
 E_{res} = E_R - i\frac{\Gamma}{2}
\end{eqnarray}

where $E_R$ represents the resonance position and $\Gamma$ is the decay width. The relation between decay 
width and lifetime $\tau$ is given by 

\begin{eqnarray}
\tau = \frac{\hbar}{\Gamma}
\end{eqnarray}

The meta-stable states are not square-integrable. They can also be seen as discrete states embedded into the continuum.
Hence
to describe  the metastable states, we  require  a  method that can simultaneously treat electron correlation and the continuum. CAP and complex scaling
\cite{comscal,comscal1,comscal2} are two 
well-known methods  used for  the  calculation  of resonance energies. Complex absorbing
potential (CAP)\cite{cap1,cap2,cap3,cap4,comscal2,cap6} along with quantum chemical methods is one of the simplest and the favoured 
approach. CAP has been implemented in many of the quantum chemical methods for the calculation of resonance states
\cite{capeom,capeom1,capeom2,capeom3a,capeom3b,capeom3c,capeom4,icdtheo1,icdtheo2,capfsa,capfsb}. 
CAP along with EOM-CCSD has been used very successfully for the study of ICD and the shape resonance phenomena.  

In the CAP approach, a one-particle potential $ -i\eta W$ is added to the physical Hamiltonian, making the original Hamiltonian complex symmetric and non-hermitian (i.e. H($\eta$) = H-i$\eta$W). 
As  a result,  we obtain  complex  eigenvalues from  the CAP  augmented  Hamiltonian (H($\eta$)). The real part denotes the resonance position, and the imaginary part  gives half of  the decay width. 
In H($\eta$) = H-i$\eta$W, $\eta$ represents the CAP strength, and W is a local positive-semidefinite one-particle operator.
With the appropriate choice of CAP, the eigenfunctions
of the complex symmetric Hamiltonian becomes square-integrable, and eigenvalues are discrete.
(H($\eta$)) is solved for various values of $\eta$.  The resonance energy is obtained by diagonalizing the complex Hamiltonian 
matrix $H(\eta)$ for multiple values of $\eta$. The  $\eta$ trajectory is obtained by plotting the real part  vs imaginary part of energy. The local minimum of trajectory is associated
with the position and half decay width of the decaying state.
\begin{eqnarray}
 \nu_i (\eta) =\eta \frac{\delta E_{i}}{\delta \eta}
\end{eqnarray}
The value of $\eta$ for which $\nu_i(\eta)$ is minimum gives the optimal CAP strength. 
We have used a box shape CAP. CAP is applied in the peripheral region so that the target remains unperturbed, yet scattered electrons are absorbed.

 In the equation of motion coupled cluster approach, the target state is generated by the action of a CI-like 
linear operator onto  the initial  reference state (closed shell coupled cluster reference state).
The wavefunction for the k$^{th}$ ionized state  $|\Psi_{k} \rangle$ is expressed as

\begin{eqnarray}
 |\Psi_{k}\rangle = R {(k)} |\Phi_{cc}\rangle
\end{eqnarray}

where R(k) is an ionization operator. The form of the linear operator for the electron ionized state can be written as 

\begin{eqnarray}
R{(k)} = \sum_{i}r_{i}(k) i+(1/2) \sum_{ija}r_{ij}^{a}(k) a^{\dagger}ji
\end{eqnarray}

The reference CC wave function, $|\Phi_{cc} \rangle$  can  be written as  

\begin{eqnarray}
|\Phi_{cc}\rangle = e^{T} |\Phi_0\rangle
\end{eqnarray}

where $ | \phi_0 \rangle $ is the  N-electron closed  shell  reference  determinant (restricted   Hartree-Fock   determinant (RHF) ) and  T  is the  cluster operator. In  the  coupled  cluster  singles  and  doubles (CCSD) approximation, the  T  operator  can  be  defined  as  follows:

\begin{eqnarray}
T=\sum_{ia}  t_{i}^{a} a_{a}^{+}  a_{i}+ 1/4 \sum_{ab} \sum_{ij} t_{ij}^{ab} a_{a}^{+} a_{b}^{+} a_{i} a_{j} 
\end{eqnarray}

The  indices a,b,...represent  the  virtual  spin orbitals  and  the  indices  i,j,.. represent the  occupied  spin  orbitals. 
In the IP-EOM-CCSD framework, the final ionized states are obtained by diagonalizing the coupled cluster
similarity transformed Hamiltonian within a $(N - 1)$ electron space.

\begin{eqnarray}
\overline{H}_{N} R (k) |\Phi_0 \rangle = \omega_k R (k) |\Phi_0 \rangle
\end{eqnarray}

%\begin{eqnarray}
%\overline{H}_{N} R (k)  = \omega_k R (k) 
%\end{eqnarray}

Where
\begin{eqnarray}
\overline{H}_{N}=e^{-T}H_{N}e^{T}-E_{cc}
\end{eqnarray}

%\begin{eqnarray}
%\overline{H}_{N}=e^{-T}H_{N}e^{T}- \langle \Phi_0| e^{-T}H_{N}e^{T} |\Phi_0 \rangle
%\end{eqnarray}

$\overline{H}_{N}$ is the similarity transformed Hamiltonian and $\omega_k$ is the energy change connected
with the ionization process. In  the IP-EOM-CCSD approach, the  matrix  is   generated   in  the  1hole   and  2hole 1particle (2h-1p) subspace. Diagonalization of
the matrix gives us ionization potential values.

In principle CAP can be implemented at the self consistent field (SCF)/ Hartree-Fock, coupled cluster (CC) or EOMCC level. Adding CAP at the coupled cluster level makes the cluster 
amplitudes complex and hence all the further calculations are complex. The N electron ground state should be unperturbed. Adding CAP at the CC or SCF level perturbs the N electron ground state itself. Then to get the correct decay width, we need to correct the N electron ground state by removing the perturbation.
% now we did not have the original energy of system (unperturbed). To get it one need to do lots of extra calculation, so we do not add CAP at SCF level. 

\begin{equation}
 |\Phi_{cc} (\eta) \rangle= e^{T(\eta)} |\Phi_{0} \rangle
\end{equation}

\begin{equation}
 \overline{H}_N (\eta)= e^{T(\eta)} H_{N}(\eta) e^{T(\eta)} - \langle \Phi_0 | e^{T(\eta)}
 H_{N}(\eta) e^{T(\eta)} | \Phi_0 \rangle 
\end{equation}

The resonance energy is obtained as 
\begin{equation}
 E_{res}(\eta)= \omega_k(\eta)-E_{cc}(\eta)-E_{cc}(\eta=0)
\end{equation}

Thus, we  lose  the  advantage  of  computing  resonance  energy  as  the  direct  energy  difference obtained  as eigenvalues  of  $\overline{H}_{N} (\eta)$   using the  IP-EOM-CCSD approach. Our previous study of
resonance \cite{pccparyya}, shows that the results are not affected when we implement CAP at the IP-EOM-CCSD level. In our calculation for the decay width, we have added 
the CAP potential in the particle-particle block of the one-particle $\overline {H}_{N}$ matrix, leaving our N electron ground state unaffected. Therefore, the  $\overline{H}_{N} (\eta)$ can  be written  as 

\begin{equation}
 \overline{H}_N (\eta)= e^{T(\eta=0)} H_{N}(\eta) e^{T(\eta=0)} - \langle \Phi_0 | e^{T(\eta=0)}
 H_{N}(\eta=0) e^{T(\eta=0)} | \Phi_0 \rangle 
\end{equation}

To generate the $\eta$ trajectory for locating the stationary point, we need to run the CAP-IP-EOM-CCSD calculations thousands of times.  We start with
$\eta=0$ and then proceed with small incremental $\eta$ values. 
Since IP-EOM-CCSD scales as $N^{6}$, it makes our calculations computationally intensive. Convergence of the equations for various $\eta$ values may be difficult as we are interested in the inner valence state, and the presence of metal ion may add to the problem. Hence we have used the full diagonalization of the matrix using BLAS routines. The dimension of the matrix usually 
is $NH+NH*NH*NP$ in a given basis for a system where NH and NP represent the number of occupied  and unoccupied orbitals.

\section{Results and discussion}\label{sec4}

This paper has studied the decay mechanism of the microsolvated clusters of small cations like Li$^{+}$, Na$^{+}$, K$^{+}$ 
with water. The bond distance, different environments and number of neighbours play an important role in the decay process.
So, we have studied the behaviour of the decay width in microsolvated clusters as  a function of the following parameters:

 a) Effect of different molecular environments on decay width for ETMD: We have studied the decay of the Lithium 1s state in Li$^{+}$-water and Li$^{+}$-ammonia.  
 Ammonia resembles the amino group found in bio-molecules and it is iso-electronic with water also. 
 This is the main reason to choose these systems to study the effect of different environments on the ETMD process. 
b) To check the impact of an increasing number of surrounding molecules on the decay width for ETMD: We have studied the decay of the Li$^{+}$ 1s state in Li$^{+}$-(H$_{2}$O)$_n$; {n=1,3}.
c) To see the effect of bond distance on the decay width for ICD: We have studied the decay of the 2s state of Na$^{+}$  in 
Na$^{+}$-H$_{2}$O  at various bond lengths. The distance between sodium and oxygen is varied. 
d) To study the effect of polarization on the decay width,  we have studied  Li$^{+}$-water and Na$^{+}$-water systems in the gaseous and liquid phases.
We have used the PCM model \cite{pcm} to study the liquid phase.   

We have studied ICD of the Sodium 2s state in
Na$^{+}$-(H$_{2}$O)$_{n}$; n=1,2 and
Auger decay of potassium 2s and 2p states in K$^{+}$-(H$_{2}$O). The details of the geometries used in this paper are available 
in the supporting information along with the basis set and method used for the geometry optimization.
Geometries were optimized using the Gaussian09 \cite{g09} software package. For the rest of the calculations, the codes used are homegrown.

\begin{table}
	\centering
%\caption{{Energy and decay width of the Li 1s state in $Li^{+}-H_{2}O$ in different basis sets }}
\caption{{Effect of variation of basis set on decay width of the Li 1s state in Li$^{+}$-H$_{2}$O for ETMD process}}
%\label{ tab:li-H_{2}O }
\begin{tabular}{|c|c|c|c|}
\hline 

\bf{Basis} & \bf{Energy (eV)} & \bf{Width(meV)} & \bf{lifetime(fs)} \\ 
\hline 
aug-cc-pVDZ      &  72.36   & 11 & 60 \\
\hline
aug-cc-pVDZ+F(O) &  72.32  & 12  & 56 \\
\hline
aug-cc-pVTZ      &  71.89  & 6.6 & 98 \\
\hline
aug-cc-pVTZ+F(O) &  72.16  & 7  & 96 \\            
%& & \\
\hline
\end{tabular}
\label{tab1}
\end{table}

%\subsection*{Decay process for the  1s ionized  state of Li$^{+}$ }
\subsection*{Choice of basis set for ETMD process of 1s state of Li$^{+}$ in Li$^{+}$-H$_2$O}

 We have studied Li$^{+}$-water in four different basis sets. Basis-A is an aug-cc-pVDZ basis set \cite{ccpvdz}.
 Basis-B is
constructed by adding an extra Rydberg type  f function on the oxygen atom of the water molecule in Basis-A. In the Li$^{+}$-water system, after ionizing 
the electron from the 1s orbital of lithium, an electron will be transferred from oxygen to fill the 1s vacancy created on lithium. The excess energy will be used to knock out a secondary electron from oxygen. Thus, it is important to have Rydberg type function on oxygen to get the continuum effect. The  exponent
of the f function are constructed according to the
method of Kaufmann et al \cite{Kaufmann}. Basis-C is an aug-cc-pVTZ basis set \cite{ccpvtz}. 
Basis-D is constructed using  the aug-cc-pVTZ basis set on lithium and oxygen atoms  and the cc-pVTZ basis set
for the hydrogen atom. In basis-D, an extra Rydberg f function  is added to the  oxygen atom similar to the  basis-B. 
The CAP box side lengths are chosen to be $C_{x}$ = $C_{y}$ = $\delta c$ and  $C_{z}$ = $\delta c + R/2$ a.u. 
The $\delta c$ value is chosen to be 5.0 a.u for the aug-cc-pVDZ basis set and 6 a.u for the aug-cc-pVTZ basis set.
Table 1 reports the resonance position and decay width (lifetime) for the Lithium 1s 
state in all four basis sets. The triple zeta (TZ) quality basis reduces the decay position by 0.5 eV. The addition of the Rydberg f 
function on oxygen has minimal effect on the decay position (i.e. ionization potential) and the decay width in double zeta and triple zeta basis sets. 
For the Li$^+$-water dimer and trimer study, we have used basis-A (i.e. aug-cc-pVDZ basis \cite{ccpvdz}) as the method scales $N^{6}$, making it computationally expensive with the higher basis set.

%\subsection*{Effect of different molecualer environment on ETMD process of  1s ionized  state of Li in Li$^{+}$-water and Li$^{+}$-ammonia }
\subsection*{Effect of different molecular environments on the ETMD process}
To see the impact of different molecular environments on the decay of the lithium 1s state, we have chosen  Li$^{+}$-NH$_{3}$ and Li$^{+}$-H$_{2}$O 
as study systems. Since they are iso-electronic systems, they are relevant systems  
to study the effect of different molecular environments on the decay width of lithium 1s state for the ETMD process. 
We have used the aug-cc-pVTZ + Rydberg 1f function on oxygen and nitrogen. 
%The CAP box size used for both the systems is $C_{x}$ = $C_{y}$ = $\delta c$ and  $C_{z}$ = $\delta c + R/2$ a.u.% 
The $\delta c$ value chosen to be 6.0 a.u.  We have presented our results in Table 2.

\begin{figure}
 \centering
 \includegraphics[width=0.75\textwidth]{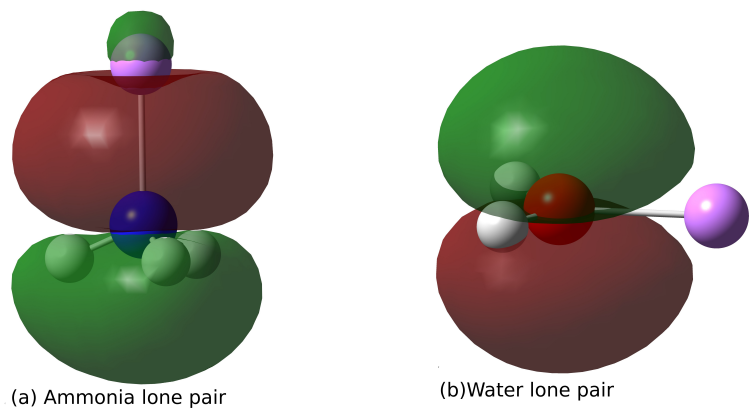}
  \caption{ a) Orientation of lone pair in Li$^{+}$-NH$_3$ is toward Li atom, making electron transfer easier, hence a shorter lifetime. b) The direction of lone pair in Li$^{+}$-H$_2$O is perpendicular to the molecular plane, making electron transfer difficult, thus more significant lifetime. Above picture is generated for isosurface value 0.02 e$\mathring{A^{-3}}$. Atom colour code Pink: Lithium, Blue: Nitrogen, Red: Oxygen, and White: Hydrogen.}
 \label{fig:lithium-ammonia-water}
\end{figure}

The ionization potential of the Li 1s state is 71.89 eV and 71.18 eV in Li$^{+}$-H$_{2}$O  and Li$^{+}$-NH$_{3}$, respectively.
The decay widths are 7 meV and 8 meV,
respectively. The decay is faster in Li$^{+}$-NH$_{3}$  (lifetime of 81 fs) than the Li$^{+}$-H$_{2}$O (lifetime of 96 fs). 
It means that the transfer of the electron to the 1s vacant position of the Li atom is faster in the case of ammonia than water. It can be explained on the following basis 
a.) Electronegativity: Oxygen is a more electronegative atom than nitrogen, making the electron transfer slower in case of water than ammonia.
b.) Position of lone pair electron: The position of the lone pair electron in the  Li$^{+}$-NH$_{3}$  is between Li$^+$ and nitrogen. In  Li$^{+}$-H$_{2}$O, the position of lone pair is perpendicular to  Li$^{+}$-H$_{2}$O molecular plane (not in between Li$^+$ and oxygen); see figure-2. Because of the directional nature of the p-orbital and lone pair's orientation toward lithium, electron transfer is much faster in case of ammonia than water. Orbitals of figure-2 are generated by Gaussian09 \cite{g09} software package using density functional theory with B3LYP functional \cite{B3,LYP,b3lyp,VWN} and aug-cc-pVTZ basis set \cite{ccpvtz}. 
The error bar of the IP-EOM-CCSD method is larger than 1meV. However, the trend should remain the same. To confirm this, we have calculated the ionization potential (IP) for both systems using CCSD(T) method. The IP values are 71.81 eV and 71.18 eV for Li$^{+}$-H$_{2}$O and Li$^{+}$-NH$_{3}$ with partial inclusion of triples, respectively. We hope that the qualitative trend for decay width will be similar. Hence concluding that the  Li$^{+}$-NH$_{3}$ decays faster than the Li$^{+}$-H$_{2}$O should remain the same.

\begin{table}
\centering
%\caption {Li 1s decay in $Li^{+}-H_{2}O$ and $Li^{+}-NH_{3} $ in aug-cc-pVTZ+1F basis}
\caption {Effect of different molecular surroundings on ETMD process for Li 1s state in Li$^{+}$-H$_{2}$O and Li$^{+}$-NH$_{3}$ using aug-cc-pVTZ+1f basis set.}
\label {tab:Li-NH_{3}}
\begin{tabular}{|c|c|c|c|}
\hline
\bf{System} & \bf{Energy (eV)} & \bf{Width (meV)} & \bf{lifetime(fs)}  \\
\hline
$Li^{+}-H_{2}O$  & 71.89  & 7 & 96 \\
\hline
$Li^{+}-NH_{3}$ &  71.18 & 8 & 81 \\
\hline

\end{tabular}
\label{tab2}
\end {table}

\subsection*{Li$^+$-water clusters: Effect of the increasing number of water molecules in surrounding on the ETMD process}
We have used the aug-cc-pVDZ basis set for the Li$^+$-water dimer and trimer. 
The CAP box size used for the dimer is $C_{x}$ = 8 a.u; $C_{y}$ = $C_{z}$ = 5 a.u. and $C_{x}$ = $C_{y}$ = 10 a.u; $C_{z}$ = 5 a.u for the trimer. The Li$^+$ ion is a good example to study the ETMD process since it has only core electrons. Therefore, ICD and Auger decay process can not take place in this system. 

If the  Li$^{+}$-water cluster is  ionized  by removing an  electron from the 1s  subshell  of the Li$^{+}$ ion, then the molecule  
relaxes  via the ETMD  process.  The ETMD  process of Li$^{2+}$(1s$^{-1}$2s$^{-1}$) state  in Li$^{+}$-water  clusters can  be  described  as 
follows. After removing  an  electron  from  the 1s orbital of  Li$^{+}$ ion in Li$^{+}$-H$_{2}$O (i.e. formation of Li$^{2+}$-H$_{2}$O), the  1s vacancy of  Li$^{2+}$(1s$^{-1}$2s$^{-1}$) ion  is  filled  up  by a 2p 
outer valence  electron of the oxygen atom  of  one  of  the  water  molecules. Then  the  released  energy  emits  a  secondary 
outer valence  electron from  the  same  water  molecule or  another  water  molecule. Therefore,  the  final  state  (Li$^{+}$-H${_2}$O$^{2+}$ or
H${_2}$O$^{+}$-Li$^{+}$-H${_2}$O$^{+}$) of the ETMD  process  is a double  ionized  state (with respect to our initial system, i.e. Li$^{+}$-H$_{2}$O). The  Li$^{+}$-H${_2}$O$^{2+}$ final  state is produced  via an ETMD(2)  process, where  both  
the  positive  charges are present  on  one  water molecule. The  
H${_2}$O$^{+}$-Li$^{+}$-H${_2}$O$^{+}$ double  ionized final  state (with respect to our initial system, i.e. Li$^{+}$-H$_{2}$O) is produced  via  ETMD(3). The ETMD  channel  is  open  for  the 1s ionized 
state  of  Li$^{+}$ ion  because  the  energy  of the Li$^{2+}$(1s$^{-1}$2s$^{-1}$) state  lies  above  the  double  ionized  final  states (i.e.  Li$^{+}$-H${_2}$O$^{2+}$ or
H${_2}$O$^{+}$-Li$^{+}$-H${_2}$O$^{+}$). Thus, the positively charged water molecules 
will repel the positively charged Lithium ion leading to a Coulomb explosion.
The different variants of ETMD (i.e. ETMD(2) or ETMD(3)) may be possible with an increasing number of water molecules surrounding the Li$^+$ ion.
The  1s ionization  energy  of the Li$^{+}$ ion in Li$^{+}$-water cluster  varies  from 72.35 to  67.45 eV depending  on  the  number 
of  water 
molecules  present  in  the  surroundings of  Li$^{+}$ ion.  Here, we have calculated  the lifetime  of  1s ionized 
state  of  Li$^{+}$ ion in various Li$^{+}$-water clusters. 

In figure-3, we have plotted the  decay values for the Li$^{+}$-(H$_{2}$O)$_{n}$;n=1,3 system.  
In this case, we have used the aug-cc-pVDZ basis set for the study. 
As we move from the monomer to the trimer, the decay position reduces from 72.35 eV to 67.45 eV.  On the other hand, the decay width increases from 11 meV to 63 meV. There are two factors that can affect the decay width: first, the  bond distance between the Li$^+$ and H$_2$O molecules and second, the number of surrounding water molecules. The bond length does not seem to have much effect as we move from the monomer (1.867 \AA) to the dimer (1.86 \AA). Therefore,  the  number  of  decay  channels  play a  significant  role in  increasing  the  decay  width  as  we move from  monomer  to  dimer. 
From figure-3, we have noticed that the decay  width  increases  nonlinearly. The  possible  reason  for the nonlinear  growth  of the decay  width is that the number of decay channels increases nonlinearly with an increasing number of water molecules surrounding the Li$^{+}$ ion.
Cederbaum and M\"{u}ller \cite{etmdhbond} have studied Li$^{+}$-H$_{2}$O with up to five water molecules using a perturbation theory ansatz with SCF integrals. They estimated lifetimes in the range of 100-20 fs. Our results give the decay time  in the range from 60 fs to 10 fs from monomer to trimer in the aug-cc-pVDZ basis \cite{ccpvdz}. In the aug-cc-pVTZ basis set \cite{ccpvtz}, we obtained a lifetime of 98 fs which is in good agreement with the Cederbaum and M\"{u}ller. See reference \cite{imke2} for details of the IP and DIP spectra of the Li$^{+}$-(H$_{2}$O)$_{n}$ complex.

%\begin{table}
% \centering
%%%\caption{{Energy and decay width of the Li 1s state in $Li^{+}-H_{2}O_{n=1,3}$ in aug-cc-pVDZ basis}}
%%% \caption{{Effect of increase in number of surrounding molecules on Li 1s state in Li$^{+}$-H$_{2}$O$_n$;n=1,3 for ETMD process in aug-cc-pVDZ basis}}

%%%\label{ tab:li-H_{2}O }
% \begin{tabular}{|c|c|c|c|}
% \hline
% System &  Energy (eV) & Width(meV) & lifetime(fs)\\
% \hline
% $Li^{+}-H_{2}O$  &  72.35 & 11 & 60 \\
% \hline
% $Li^{+}-(H_{2}O)_{2}$ &69.17    &  40 & 16    \\
% \hline
% $Li^{+}-(H_{2}O)_{3}$ & 67.45    & 63 & 10 \\
% \hline

 %\end{tabular}
 %\label{tab3}
%\end{table}

\begin{figure}
    \centering
 \includegraphics[width=0.75\textwidth]{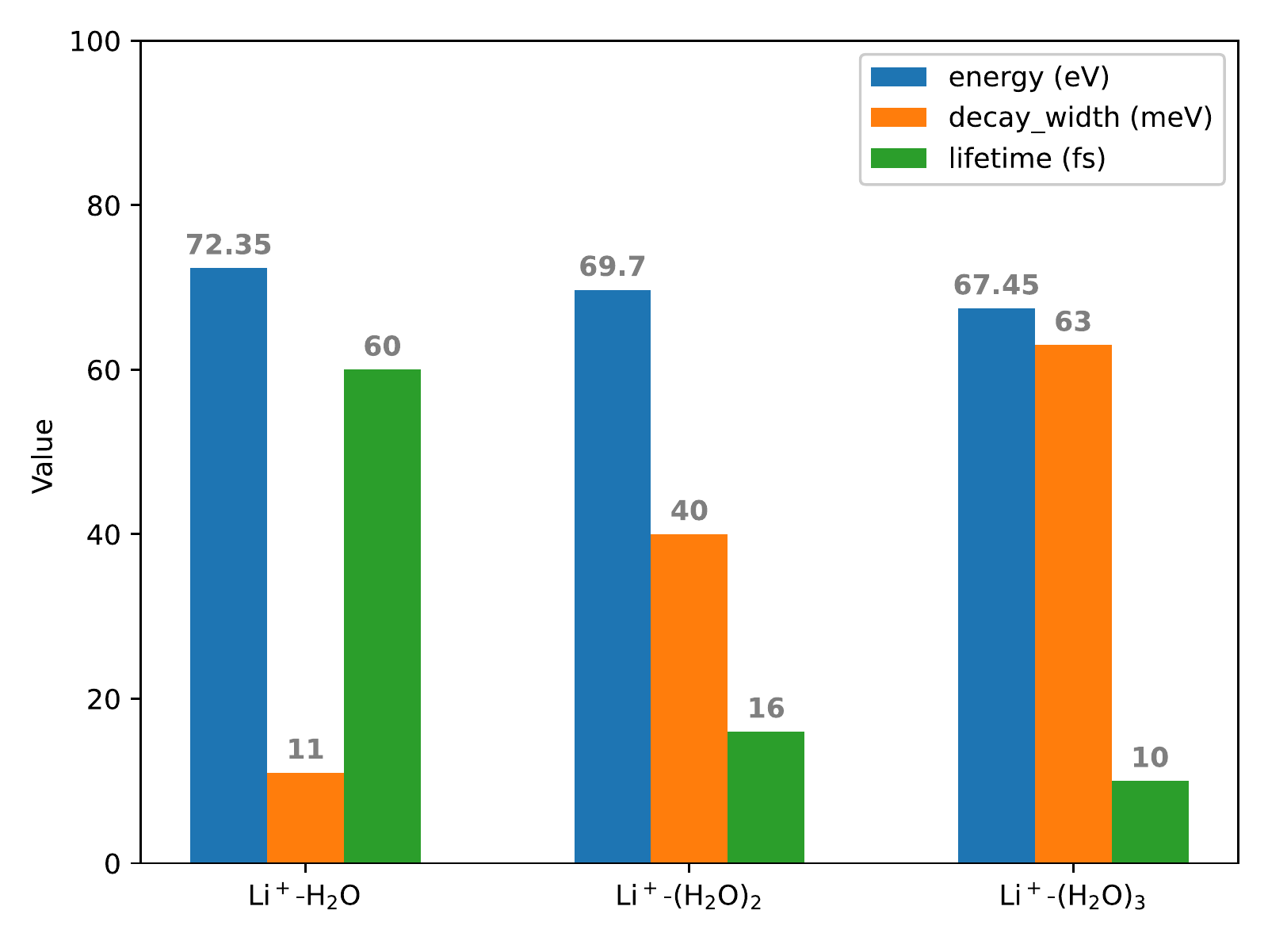}
%\vspace {-15cm}
    \caption{ Effect of the increased  number of surrounding water molecules on ETMD process using an aug-cc-pVDZ basis set. }
    \label{fig:lithium-water}
\end{figure}

%\subsection*{Intermolecular coulombic decay process for the 2s ionized state of  Na atom  in Na$^{+}$-water cluster } 
\subsection*{Na$^{+}$-water cluster: Effect of distance and increasing number of water molecules on the ICD process} 
The  2s ionized  state  of the Na  atom in the Na$^{+}$-H$_{2}$O system  can  relax  via  the ICD process. The  ICD  process of the 2s ionized  state  of the Na atom in the  
 Na$^{+}$-H$_{2}$O system  can  be  described  as follows: after  removing an  electron  from  the 2s subshell  of  Na atom,  the  2s 
 vacancy  of the Na atom in the Na$^{2+}$-H$_{2}$O is  filled  up  by a 2p  outer valence  electron  of the Na atom. Then 
 the  released  energy is  transferred to  the  neighbouring  H$_2$O molecule  which  emits  a  
 secondary  electron.  Therefore, the  final  state  of the ICD  process  is  characterized  by 
the Na$^{2+}$ (2p$^{-1}$3s$^{-1}$) O$^{+}$H$_2$ (2p$^{-1}$) triple  ionized  state. Energetically, the ICD  process  is  possible  
in the Na$^{+}$-H$_{2}$O system because  the  energy  of the 2s  ionized state  of the Na  atom  lies  above  the  energy  of
the Na$^{2+}$ (2p$^{-1}$3s$^{-1}$) O$^{+}$H$_2$(2p$^{-1}$) final  state.
 
 The decay of the Sodium 2s state in the Na$^{+}$-H$_{2}$O system is studied in a modified maug-cc-pV(T+d)Z \cite{minmalaug} basis set and augmented by 3s3p1d functions for Oxygen atom only(taken from the basis set exchange library \cite{esml}, then modified) for sodium and 
 oxygen and cc-pVDZ \cite{ccpvdz} for hydrogen. The detailed basis set used for the system is described in supporting information. 
 To see the effect of bond length on the decay width, we have studied the Na$^+$-water system for various bond lengths, i.e. 2.24988 $\mathring{A}$ to 5.0 $\mathring{A}$ between  the sodium  and the oxygen  atom. The geometry was optimized using the CCSD method in the aug-cc-pVDZ basis \cite{ccpvdz} and the bond distance between sodium and oxygen was found to be 2.24988 \AA. Thus, we have used this bond distance.  The CAP box size used in our calculation
is $C_{x}$ = 8.7 ,$C_{y}$ = 6.5 and   $C_{z}$ = 5.0 a.u. Fig 4 summarizes the results for various bond lengths. We know from equation 2 that the decay width and lifetime of a temporary bound state (TBS) are inversely proportional to each other. We observe that the lifetime increases rapidly as bond length increases or the decay width rapidly decreases with an increase in the bond length. A sharp change in decay width from 251 meV (2.6 fs) at 2.24988 $\mathring{A}$ to 142 meV (4.6 fs) at 2.29 $\mathring{A}$ is observed. Then it saturates at 33 meV (20 fs) at 5.0$\mathring{A}$. It may become bound with a further increase in the bond length. We also report the Fano-ADC \cite{microsolv} results at 2.30 \AA. The basis 
set used in that calculation is cc-pCVTZ+2s2p2d1f KBJ \cite{Kaufmann} for sodium and oxygen and  cc-pVTZ+1s1p1d  KBJ  \cite{Kaufmann} for hydrogen. 
The lifetime of 7 fs was reported using the Fano-ADC method. The experimental \cite{Nawaterexpt} value of the decay time of the Na 2s state is 3.1 fs. 
Our results predict the decay time to be 2.6 fs at the equilibrium bond length which shows good agreement with the experimental value.

\begin{figure}
 \centering
 \includegraphics[width=0.9\textwidth]{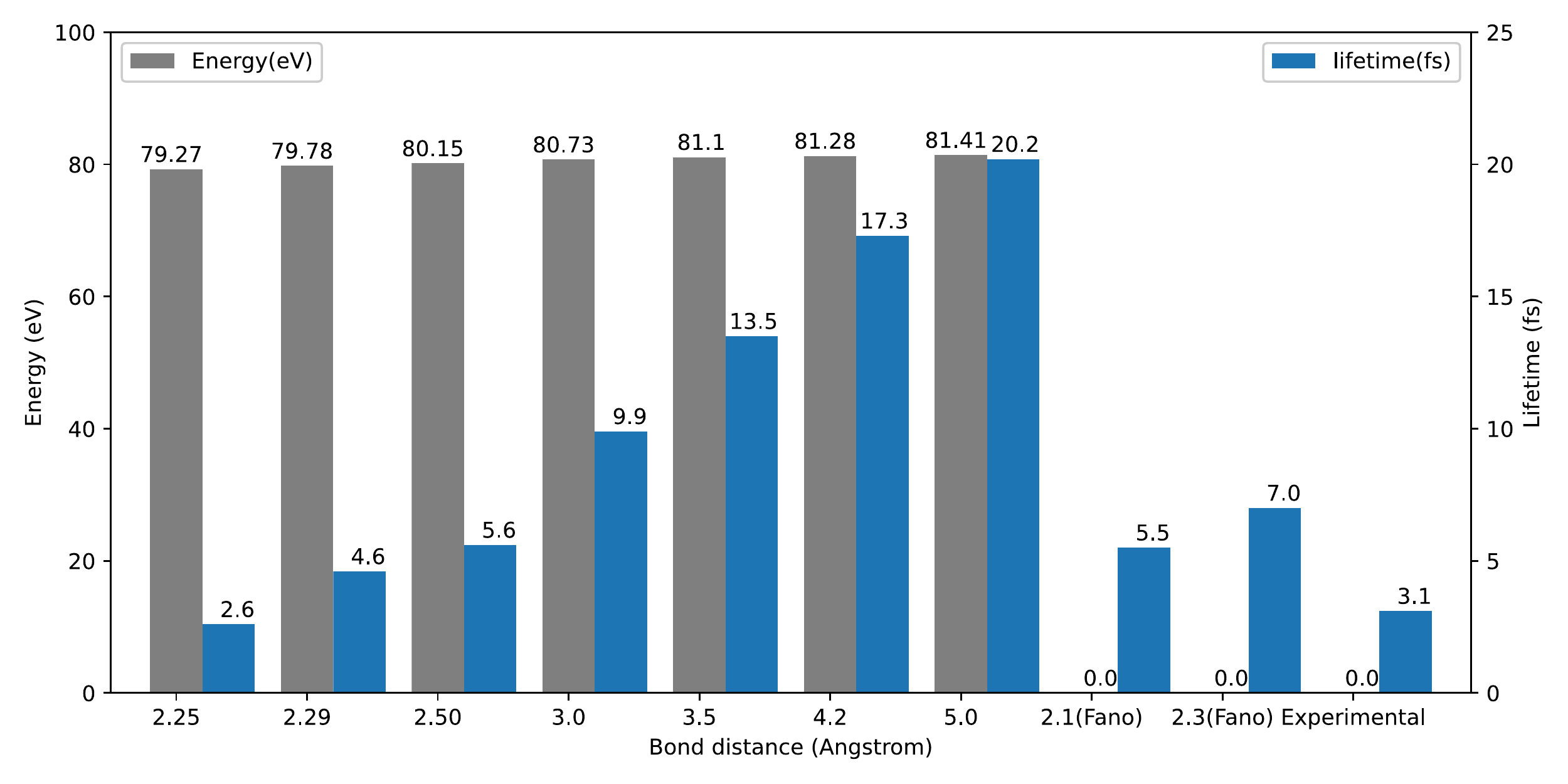}
%\includegraphics[width=0.75\textwidth]{Na-water_DIS.eps}
%\vspace {-15cm}
    \caption{ Effect of bond length on ICD process for the 2s state of Na in Na$^+$-H$_2$O using maug-cc-pV(T+d)Z basis set.}
    \label{fig:sodium-water}
\end{figure}

\begin{table}
\centering
\caption{Different decay process in different systems: Auger decay in K$^{+}$-H$_{2}$O, ICD in Na$^{+}$-(H$_{2}$O)$_n$ and ETMD in Li$^{+}$-(H$_{2}$O)$_{n}$ (where n=1,2).}
%\label{Li-XN}

\begin{tabular}{|c|c|c|c|}
\hline
\bf{Basis} & \bf{System} & \bf{Energy (eV)} & \bf{Width in meV (fs)} \\
\hline
maug-cc-pV(T+d)Z$^@$   & Li$^{+}$-H$_{2}$O  & 71.81 & 7.64 (86) \\
\hline
maug-cc-pV(T+d)Z$^@$  & Li$^{+}$-(H$_{2}$O)$_{2}$ & 69.03    & 16.12 (40.8)    \\
\hline
maug-cc-pV(T+d)Z$^{**}$  & Na$^{+}$-H$_{2}$O  & 79.49 & 129 (5.1) \\
\hline
maug-cc-pV(T+d)Z$^{**}$  & Na$^{+}$-(H$_{2}$O)$_{2}$ &78.19   & 305 (2.1)   \\
\hline
 aug-cc-pVDZ-X2C   & K$^{+}$-H$_{2}$O (2s) &  396 & 423 (1.5)\\
                   &                       &      & 278 (2.4)$^*$   \\
\hline
 aug-cc-pVDZ-X2C & K$^{+}$-H$_{2}$O (2p) & 315  & 74.86 (8.8)   \\
                 &                       &      & 246 (2.7)$^*$   \\
                 
\hline
\end{tabular}\\
* represents the 2nd decay value that we have observed.\\
** represents the maug-cc-pV(T+d)Z + 3s3p1d on O atom in Spherical basis.\\
@ represents the maug-cc-pV(T+d)Z + 3s3p1d on O atom in Cartesian basis. 
\label{tab5}
\end {table}

To see the effect of the increased number of surrounding molecules on the decay width in the ICD process, we also studied the  Na$^{+}$-water dimer. The  optimized  geometry was obtained  using the  
B3LYP functional \cite{B3,LYP,b3lyp,VWN} and the 6-311++g(3d,2p) basis set \cite{Na2Ar6311g} in the Gaussian09 \cite{g09} software package. 
The CAP box size used for the Na$^{+}$-water system in our calculation is $C_{x}$ = 12 ,$C_{y}$ = 7 and   $C_{z}$ = 7 a.u. 
We have used a spherical basis set for the Na$^{+}$-water dimer due to scaling of the CC equations (i.e. in a cartesian basis, the Na$^{+}$-water dimer is computationally very expensive). To compare Na$^+$-(H$_2$O)$_2$ with  Na$^+$-H$_2$O, we again run Na$^+$-H$_2$O in a spherical basis set.  The bond distance between Sodium and Oxygen is 2.2453 $\mathring{A}$ in  Na$^+$-(H$_2$O)$_2$  and 2.24988 $\mathring{A}$ in Na$^+$-H$_2$O. The decay width of the 2s state of Sodium in Na$^+$-(H$_2$O)$_2$  is 305 meV (2.1 fs) compared to 129 meV (5.1 fs) in Na$^+$-H$_2$O.

\subsection*{Auger Decay process  for  the 2s and 2p  ionized  states of  K  in  K$^{+}$-water}

The  2s and 2p  ionized states of the K atom  in the  K$^{+}$-H$_2$O can  relax  via the Auger process.  In  our calculations,
we  have  used  the  optimized  geometry for K$^{+}$-H$_2$O  obtained  using the B3LYP functional \cite{B3,LYP,b3lyp,VWN} and the 6-311++g(3d,2p) basis set \cite{K6311g} in the Gaussian09 \cite{g09} software package.  
For  the  computation of the decay  width,  we  have  employed the 
aug-cc-pVDZ-X2C basis set \cite{x2cbasis} for potassium, aug-cc-pVDZ \cite{ccpvdz} for oxygen and the cc-pVDZ basis set \cite{ccpvdz} for hydrogen. The results for
the  2s and  2p  ionized  states  are  presented  in Table-3 along with Li$^+$-(H$_2$O)$_n$ and Na$^+$-(H$_2$O)$_n$; n=1,2. Here, for the Li$^+$-(H$_2$O)$_n$; n=1,2, we have used the maug-cc-pV(T+d)Z \cite{minmalaug}+ 3s3p1d basis set on Oxygen while for Li we have used maug-cc-pV(T+d)Z basis set\cite{minmalaug} to maintain consistency.
The CAP box size used for K$^+$-water system in our calculation is $C_{x}$ = 7 ,$C_{y}$ = 4 and   $C_{z}$ = 4 a.u. In the case of the K$^+$-H$_2$O system, we observed two stationary points
on the $\eta$ trajectory indicating decay  through a cascade  mechanism. Experimentally, a similar kind of two stationary points (cascade decay type effect) was observed \cite{kcl-water}. The $\eta$ trajectory shows only one stationary point for the other two systems (Li$^+$-H$_2$O, Na$^+$-H$_2$O). 

The Auger  process  of the (2s, 2p)  ionized  state  of the  K  atom in  K$^{+}$-H$_2$O (4s$^{-1}$) can be rationalized  as  follows:
After  removing  an  electron  from the 2s or  2p  subshell  of the K  atom in  K$^{+}$-H$_2$O (formation of K$^{2+}$-H$_2$O)(2s$^{-1}$ 4s$^{-1}$ or 2p$^{-1}$ 4s$^{-1}$), the  2s or  2p  vacancy  is  filled  up
by a 3p or  3s  outer valence electron  of the  K  atom. Then  the  released energy is used to knock out  another  secondary  
outer valence electron  from the 3p or  3s  subshell of the K atom (formation of K$^{3+}$-H$_2$O). This two-hole state ( K$^{3+}$(3p$^{-2}$ 4s$^{-1}$)H$_2$O or  K$^{3+}$(3p$^{-1}$ 3s$^{-1}$ 4s$^{-1}$)H$_2$O or 
K$^{3+}$(3s$^{-2}$ 4s$^{-1}$)H$_2$O) (which is with respect to our initial system  K$^{+}$-H$_2$O) is unstable and further relax via another decay process which is a three-hole state. 
%Therefore, the  final  state  of  Auger  process  is  identified by K$^{3+}$(3p$^{-2}$ 4s$^{-1}$)H$_2$O,  K$^{3+}$(3p$^{-1}$ 3s$^{-1}$ 4s$^{-1}$)H$_2$O, K$^{3+}$(3s$^{-2}$ 4s$^{-1}$)H$_2$O triple  ionized  final  states. The  energies  of  2s as well as  2p  ionized  state  of  K  atom  in K$^{+}$-H$_2$O cluster  are  396 eV  and  315  eV  respectively. 
Energetically the Auger  process  will be viable if  the  energy  of  the  2s or  2p  ionized  state  of the K  atom  lies  above  the  triple  ionized final  states. The calculation of the three-hole state is beyond the scope of this paper.
 
 The Auger  decay  width  for  the 2p  ionized  state  is 75 meV (i.e. 8.8 fs) which undergoes further decay with a decay width of
 246 meV corresponding to  a  lifetime  of  2.7 fs. Similarly, the  Auger decay  width  for the  2s ionized  state is  423 meV with a lifetime of 
1.5 fs which undergoes further decay with a decay width of 278 meV corresponding to 
a  lifetime  of 2.4 fs.  
Pokapanich et al. \cite{watersol1} have studied the Auger decay in 
potassium  chloride surrounded by water molecules.

%\subsection*{Polarization effect on ETMD and ICD}
\subsection*{Polarized Surrounding effect on ETMD and ICD}

We have studied the Li$^{+}$-water and Na$^{+}$-water in the gaseous and aqueous medium to see the effect of a polarized surrounding 
on the decay width in the ETMD and ICD processes, respectively.
We have used the PCM model \cite{pcm} for the aqueous phase, where water is the solvent. The bond distance and
CAP box size were kept identical for the gaseous and aqueous phases. 
The results are presented in Table 4. For the Li$^{+}$-water 1s state, the decay position remains almost identical; however, the decay width changes from 
12 meV (56 fs) for the gaseous medium to 9.7 meV (67 fs) for the aqueous medium. The decay is slower in the aqueous  medium compared to the gaseous medium. A similar trend was
observed for the Na$^{+}$-water's 2s state and the decay width changes from 143 meV (4.6) for the gaseous medium to 108 meV (6 fs) for the aqueous medium. The slow decay in the aqueous medium is due to the polarization provided by the medium which makes the ionized state more
stable than the gaseous medium.

\begin {table}
\centering
%\caption{decay of inner valence in $X^{+}-H_{2}O$ }
\caption{Polarization effect on decay width in ETMD and ICD processes for X$^{+}$-H$_{2}$O system (X = Li, Na). }
%\label{Li-Xpcm}
\begin{tabular}{|c|c|c|c|c|}
\hline
\bf{Basis} & \bf{System} & \bf{medium} & \bf{Energy (eV)} & \bf{Width in meV(fs)} \\
\hline
 aug-cc-pVDZ +F(O) &$Li^{+}-H_{2}O$  & GAS& 72.36  & 12 (56) \\
\hline
 aug-cc-pVDZ+F(O) &$Li^{+}-H_{2}O$  & PCM & 72.3  & 9.7 (67) \\
\hline
m-aug-cc-PV(T+d)Z  &$Na^{+}-H_{2}O$  & GAS & 79.78 & 143 (4.6) \\
\hline
m-aug-cc-PV(T+d)Z  &$Na^{+}-H_{2}O$  & PCM & 79.73 & 108 (6.0) \\
\hline
\end{tabular} \\

\label{tab6}

\end{table}  

\section{Conclusions}\label{sec5}

This paper has used the CAP-IP-EOM-CCSD method to study the  various decay  processes in microsolvated  alkali  metal  ions i.e. Li$^{+}$, Na$^{+}$
and K$^{+}$. The CAP-IP-EOM-CCSD method is used for the first time to explore the ETMD lifetimes for the Li 1s state in Li$^{+}$-H$_{2}$O clusters.
It is observed that the decay widths are sensitive to the bond length,  surrounding atoms, medium (gas or liquid) and the number of neighboring molecules. We have studied the effect of all these parameters on the decay width of Li$^{+}$-H$_{2}$O and Na$^{+}$-H$_{2}$O clusters.

We have studied the decay of  1s, 2s, and both 2s and 2p states in Li$^{+}$, Na$^{+}$ and K$^{+}$ with water, respectively.   
The Li 1s state 
undergoes ETMD whereas the Na 2s state decays via ICD. The K 2s and 2p states undergo Auger decay.  To study the impact of different molecular environments, the 1s  ionized  state  of the Li  atom was studied 
in  Li$^{+}$-NH$_3$ and Li$^{+}$-H$_{2}$O. Since water and ammonia are isoelectronic, it will be interesting to study the effect of the environment on the  decay width. We found that decay is faster in Li$^{+}$-NH$_3$ (81 fs)  than in  Li$^{+}$-H$_{2}$O (96 fs). 
The possible explanation for this could be, first, the higher electronegativity of oxygen than nitrogen, making electron transfer more  difficult than nitrogen. Second, the location of the lone pair. In Li$^{+}$-NH$_3$, it is between Li$^{+}$ and nitrogen, whereas in Li$^{+}$-H$_{2}$O, it is perpendicular to the molecular plane. Because of the directional nature of p-orbitals and the orientation of the lone pair toward lithium, electron transfer is much faster in the case of ammonia than water (See Figure-2 for details).

%Li$^{+}$-H$_{2}$O is more polarizable than the Li$^{+}$-NH$_3$ which reduces the decay width and hence more life time.  
 
We  have  studied  the  ETMD lifetime for the 1s ionized state of the Li atom in the Li$^{+}$-(H$_{2}$O)$_{n}$ (n=1,3) system to see the 
effect of the number of water molecules on the decay. 
The  lifetime  obtained  for  the   Li$^{+}$-(H$_{2}$O)$_{n}$ system  is  60 fs, 16 fs and drops  further  to  10 fs as n increases from 1 to 3. 
 As a  characteristic  feature  of  ETMD, the  lifetime  decreases  strongly  with  an  increasing  number  of  neighbours. This is due to a nonlinear increase in the number of decay channels with an increasing number of surrounding atoms.

To see the effect of bond length on the decay width for the ICD process, we studied the  2s  ionized  state  of  Na  atom in Na$^{+}$-H$_{2}$O at various  bond lengths, i.e. 2.2489 $\mathring{A}$ to  5.0 $\mathring{A}$. We observe that as the bond
length increases, the decay width reduces and the lifetime increases from 2.6 fs at 2.24 $\mathring{A}$ to 20 fs at 
5.0 $\mathring{A}$. A similar trend was observed  using the Fano ADC method \cite{microsolv}. The authors report a lifetime of 5.5 fs at 2.21 $\mathring{A}$ and 7 fs at 2.30 $\mathring{A}$. Our results for sodium 2s state are in good
agreement with the experiment \cite{Nawaterexpt} and the Fano ADC \cite{microsolv} method.  
We have  also  investigated  the  ICD  lifetime  for  the  2s  ionized  state  of the Na  atom in the Na$^{+}$-(H$_{2}$O)$_{n}$ 
(n=1,2) systems to study the effect of increased water molecules in the surrounding on ICD.  The  computed ICD lifetime  for  the Na$^{+}$-H$_{2}$O system  is  5.1 fs, and  it  decreases  strongly 
to  2.1 fs for  the  Na$^{+}$-(H$_{2}$O)$_{2}$ system.  We have used a spherical Gaussian basis set here for monomer and dimer to have a proper 
comparison. The sensitivity of the decay width to the spherical or cartesian  basis is also observed.   
We  have investigated  the  Auger lifetime  for  the  2s and  2p  ionized  state  of the K  atom in 
the K$^{+}$-(H$_2$O) system. The  computed  Auger  lifetimes  for  the 2s and  2p  ionized states are 2.36 fs and   2.67 fs, respectively.
The $\eta$ trajectory indicates a cascade decay for the K$^{+}$-(H$_2$O)  system.  The Auger decay initiates another decay after a short-lived state leading to a more stable state. Since the Auger decay is a localized decay, we do not expect much change with bond length or number of neighbours. 

To know the polarization effect on the ICD and ETMD processes, we studied the decay of the 1s state of the Li atom and the 2s state of Na in Li$^{+}$-(H$_{2}$O) and
Na$^{+}$-(H$_{2}$O), respectively. We used the PCM model \cite{pcm} in our study. In both cases, our results show that the polarization 
stabilizes the system, i.e. decay time is increased in the liquid phase compared to the gaseous phase. 
In this paper, we studied the decay width of alkali metal ions as a function of  different molecular environments, increase of surrounding molecules in a system, bond distance, basis set and polarization of the medium.

\begin{acknowledgement}

NV acknowledges financial support from Department of Science and Technology. RK acknowledges financial support from Council of 
scientific and Industrial Research. Authors acknowledge the computational facility at the CSIR-National Chemical Laboratory.
We dedicate this paper to Prof. Sourav Pal on his 65th birthday.

\end{acknowledgement}

\begin{suppinfo}

The data that support the findings of this study are available as supporting information submitted to ACS. Any further data will be available on request from corresponding authors.

\end{suppinfo}

\bibliography{ct-2021-01036p.bib}

\providecommand{\latin}[1]{#1}
\makeatletter
\providecommand{\doi}
  {\begingroup\let\do\@makeother\dospecials
  \catcode`\{=1 \catcode`\}=2 \doi@aux}
\providecommand{\doi@aux}[1]{\endgroup\texttt{#1}}
\makeatother
\providecommand*\mcitethebibliography{\thebibliography}
\csname @ifundefined\endcsname{endmcitethebibliography}
  {\let\endmcitethebibliography\endthebibliography}{}
\begin{mcitethebibliography}{80}
\providecommand*\natexlab[1]{#1}
\providecommand*\mciteSetBstSublistMode[1]{}
\providecommand*\mciteSetBstMaxWidthForm[2]{}
\providecommand*\mciteBstWouldAddEndPuncttrue
  {\def\EndOfBibitem{\unskip.}}
\providecommand*\mciteBstWouldAddEndPunctfalse
  {\let\EndOfBibitem\relax}
\providecommand*\mciteSetBstMidEndSepPunct[3]{}
\providecommand*\mciteSetBstSublistLabelBeginEnd[3]{}
\providecommand*\EndOfBibitem{}
\mciteSetBstSublistMode{f}
\mciteSetBstMaxWidthForm{subitem}{(\alph{mcitesubitemcount})}
\mciteSetBstSublistLabelBeginEnd
  {\mcitemaxwidthsubitemform\space}
  {\relax}
  {\relax}

\bibitem[Auger(1925)]{Auger}
Auger,~P. Sur l' effet Photo\'{e}lectrique Compos\'{e}. \emph{J. Phys. Radium}
  \textbf{1925}, \emph{6}, 205--208\relax
\mciteBstWouldAddEndPuncttrue
\mciteSetBstMidEndSepPunct{\mcitedefaultmidpunct}
{\mcitedefaultendpunct}{\mcitedefaultseppunct}\relax
\EndOfBibitem
\bibitem[Cederbaum \latin{et~al.}(1997)Cederbaum, Zobeley, and Tarantelli]{ICD}
Cederbaum,~L.~S.; Zobeley,~J.; Tarantelli,~F. Giant Intermolecular Decay and
  Fragmentation of Clusters. \emph{Phys. Rev. Lett.} \textbf{1997}, \emph{79},
  4778--4781\relax
\mciteBstWouldAddEndPuncttrue
\mciteSetBstMidEndSepPunct{\mcitedefaultmidpunct}
{\mcitedefaultendpunct}{\mcitedefaultseppunct}\relax
\EndOfBibitem
\bibitem[Santra and Cederbaum(2002)Santra, and Cederbaum]{ICD1}
Santra,~R.; Cederbaum,~L.~S. Non-Hermitian Electronic Theory and Applications
  to Clusters. \emph{Phys. Rep.} \textbf{2002}, \emph{368}, 1--117\relax
\mciteBstWouldAddEndPuncttrue
\mciteSetBstMidEndSepPunct{\mcitedefaultmidpunct}
{\mcitedefaultendpunct}{\mcitedefaultseppunct}\relax
\EndOfBibitem
\bibitem[Zobeley \latin{et~al.}(2001)Zobeley, Santra, and Cederbaum]{etmd}
Zobeley,~J.; Santra,~R.; Cederbaum,~L.~S. Electronic Decay in Weakly Bound
  Heteroclusters: Energy Transfer versus Electron Transfer. \emph{J. Chem.
  Phys.} \textbf{2001}, \emph{115}, 5076--5088\relax
\mciteBstWouldAddEndPuncttrue
\mciteSetBstMidEndSepPunct{\mcitedefaultmidpunct}
{\mcitedefaultendpunct}{\mcitedefaultseppunct}\relax
\EndOfBibitem
\bibitem[Scheit \latin{et~al.}(2004)Scheit, Averbukh, Meyer, Moiseyev, Santra,
  Sommerfeld, Zobeley, and Cederbaum]{ICDtheo}
Scheit,~S.; Averbukh,~V.; Meyer,~H.; Moiseyev,~N.; Santra,~R.; Sommerfeld,~T.;
  Zobeley,~J.; Cederbaum,~L.~S. On the Interatomic Coulombic Decay in the Ne
  Dimer. \emph{J. Chem. Phys.} \textbf{2004}, \emph{121}, 8393--8398\relax
\mciteBstWouldAddEndPuncttrue
\mciteSetBstMidEndSepPunct{\mcitedefaultmidpunct}
{\mcitedefaultendpunct}{\mcitedefaultseppunct}\relax
\EndOfBibitem
\bibitem[Vaval and Cederbaum(2007)Vaval, and Cederbaum]{icdtheo1}
Vaval,~N.; Cederbaum,~L.~S. Ab Initio Lifetimes in the Interatomic Coulombic
  Decay of Neon Clusters Computed with Propagators. \emph{J. Chem. Phys.}
  \textbf{2007}, \emph{126}, 164110\relax
\mciteBstWouldAddEndPuncttrue
\mciteSetBstMidEndSepPunct{\mcitedefaultmidpunct}
{\mcitedefaultendpunct}{\mcitedefaultseppunct}\relax
\EndOfBibitem
\bibitem[Ehara and Sommerfeld(2012)Ehara, and Sommerfeld]{icdtheo2}
Ehara,~M.; Sommerfeld,~T. CAP/SAC-CI Method for Calculating Resonance States of
  Metastable Anions. \emph{Chem. Phys. Lett.} \textbf{2012}, \emph{537},
  107--112\relax
\mciteBstWouldAddEndPuncttrue
\mciteSetBstMidEndSepPunct{\mcitedefaultmidpunct}
{\mcitedefaultendpunct}{\mcitedefaultseppunct}\relax
\EndOfBibitem
\bibitem[Jahnke(2015)]{ICDexpt}
Jahnke,~T. Interatomic and Intermolecular Coulombic Decay: The Coming of Age
  Story. \emph{J. Phys. B: At. Mol. Opt. Phys.} \textbf{2015}, \emph{48},
  082001\relax
\mciteBstWouldAddEndPuncttrue
\mciteSetBstMidEndSepPunct{\mcitedefaultmidpunct}
{\mcitedefaultendpunct}{\mcitedefaultseppunct}\relax
\EndOfBibitem
\bibitem[Marburger \latin{et~al.}(2003)Marburger, Kugeler, Hergenhahn, and
  M\"oller]{ICDexpt1}
Marburger,~S.; Kugeler,~O.; Hergenhahn,~U.; M\"oller,~T. Experimental Evidence
  for Interatomic Coulombic Decay in Ne Clusters. \emph{Phys. Rev. Lett.}
  \textbf{2003}, \emph{90}, 203401\relax
\mciteBstWouldAddEndPuncttrue
\mciteSetBstMidEndSepPunct{\mcitedefaultmidpunct}
{\mcitedefaultendpunct}{\mcitedefaultseppunct}\relax
\EndOfBibitem
\bibitem[Jahnke \latin{et~al.}(2004)Jahnke, Czasch, Sch\"{o}ffler,
  Sch\"{o}ssler, Knapp, K\"{a}sz, Titze, Wimmer, Kreidi, Grisenti, Staudte,
  Jagutzki, Hergenhahn, Schmidt-B\"{o}cking, and D\"{o}rner]{ICDexpt2}
Jahnke,~T.; Czasch,~A.; Sch\"{o}ffler,~M.~S.; Sch\"{o}ssler,~S.; Knapp,~A.;
  K\"{a}sz,~M.; Titze,~J.; Wimmer,~C.; Kreidi,~K.; Grisenti,~R.~E.;
  Staudte,~A.; Jagutzki,~O.; Hergenhahn,~U.; Schmidt-B\"{o}cking,~H.;
  D\"{o}rner,~R. Experimental Observation of Interatomic Coulombic Decay in
  Neon Dimers. \emph{Phys. Rev. Lett.} \textbf{2004}, \emph{93}, 163401\relax
\mciteBstWouldAddEndPuncttrue
\mciteSetBstMidEndSepPunct{\mcitedefaultmidpunct}
{\mcitedefaultendpunct}{\mcitedefaultseppunct}\relax
\EndOfBibitem
\bibitem[\"{O}hrwall \latin{et~al.}(2004)\"{O}hrwall, Tchaplyguine, Lundwall,
  Feifel, Bergersen, Rander, Lindblad, Schulz, Peredkov, Barth, Marburger,
  Hergenhahn, Svensson, and Bj\"{o}rneholm]{ICDexpt3}
\"{O}hrwall,~G.; Tchaplyguine,~M.; Lundwall,~M.; Feifel,~R.; Bergersen,~H.;
  Rander,~T.; Lindblad,~A.; Schulz,~J.; Peredkov,~S.; Barth,~S.; Marburger,~S.;
  Hergenhahn,~U.; Svensson,~S.; Bj\"{o}rneholm,~O. Femtosecond Interatomic
  Coulombic Decay in Free Neon Clusters: Large Lifetime Differences between
  Surface and Bulk. \emph{Phys. Rev. Lett.} \textbf{2004}, \emph{93},
  173401\relax
\mciteBstWouldAddEndPuncttrue
\mciteSetBstMidEndSepPunct{\mcitedefaultmidpunct}
{\mcitedefaultendpunct}{\mcitedefaultseppunct}\relax
\EndOfBibitem
\bibitem[Santra \latin{et~al.}(2000)Santra, Zobeley, Cederbaum, and
  Moiseyev]{icdrare}
Santra,~R.; Zobeley,~J.; Cederbaum,~L.~S.; Moiseyev,~N. Interatomic Coulombic
  Decay in Van Der Waals Clusters and Impact of Nuclear Motion. \emph{Phys.
  Rev. Lett.} \textbf{2000}, \emph{85}, 4490--4493\relax
\mciteBstWouldAddEndPuncttrue
\mciteSetBstMidEndSepPunct{\mcitedefaultmidpunct}
{\mcitedefaultendpunct}{\mcitedefaultseppunct}\relax
\EndOfBibitem
\bibitem[Sisourat \latin{et~al.}(2010)Sisourat, Sann, Kryzhevoi, Koloren\u{c},
  Havermeier, Sturm, Jahnke, Kim, D\"{o}rner, and Cederbaum]{icdhene}
Sisourat,~N.; Sann,~H.; Kryzhevoi,~N.~V.; Koloren\u{c},~P.; Havermeier,~T.;
  Sturm,~F.; Jahnke,~T.; Kim,~H.~K.; D\"{o}rner,~R.; Cederbaum,~L.~S.
  Interatomic Electronic Decay Driven by Nuclear Motion. \emph{Phys. Rev.
  Lett.} \textbf{2010}, \emph{105}, 173401\relax
\mciteBstWouldAddEndPuncttrue
\mciteSetBstMidEndSepPunct{\mcitedefaultmidpunct}
{\mcitedefaultendpunct}{\mcitedefaultseppunct}\relax
\EndOfBibitem
\bibitem[Stoychev \latin{et~al.}(2010)Stoychev, Kuleff, and
  Cederbaum]{icdhbond}
Stoychev,~S.~D.; Kuleff,~A.~I.; Cederbaum,~L.~S. On the Intermolecular
  Coulombic Decay of Singly and Doubly Ionized States of Water Dimer. \emph{J.
  Chem. Phys.} \textbf{2010}, \emph{133}, 154307\relax
\mciteBstWouldAddEndPuncttrue
\mciteSetBstMidEndSepPunct{\mcitedefaultmidpunct}
{\mcitedefaultendpunct}{\mcitedefaultseppunct}\relax
\EndOfBibitem
\bibitem[Th\"{u}rmer \latin{et~al.}(2013)Th\"{u}rmer, Unger, Slav\'{i}\u{c}ek,
  and Winter]{watersol}
Th\"{u}rmer,~S.; Unger,~I.; Slav\'{i}\u{c}ek,~P.; Winter,~B. Relaxation of
  Electronically Excited Hydrogen Peroxide in Liquid Water: Insights from
  Auger-Electron Emission. \emph{J. Phys. Chem. C} \textbf{2013}, \emph{117},
  22268--22275\relax
\mciteBstWouldAddEndPuncttrue
\mciteSetBstMidEndSepPunct{\mcitedefaultmidpunct}
{\mcitedefaultendpunct}{\mcitedefaultseppunct}\relax
\EndOfBibitem
\bibitem[Pokapanich \latin{et~al.}(2011)Pokapanich, Kryzhevoi, Ottosson,
  Svensson, Cederbaum, \"{O}hrwall, and Bj\"{o}rneholm]{watersol1}
Pokapanich,~W.; Kryzhevoi,~N.~V.; Ottosson,~N.; Svensson,~S.; Cederbaum,~L.~S.;
  \"{O}hrwall,~G.; Bj\"{o}rneholm,~O. Ionic-Charge Dependence of the
  Intermolecular Coulombic Decay Time-Scale for Aqueous Ions Probed by the
  Core-Hole Clock. \emph{J. Am. Chem. Soc.} \textbf{2011}, \emph{133},
  13430--13436\relax
\mciteBstWouldAddEndPuncttrue
\mciteSetBstMidEndSepPunct{\mcitedefaultmidpunct}
{\mcitedefaultendpunct}{\mcitedefaultseppunct}\relax
\EndOfBibitem
\bibitem[Fasshauer \latin{et~al.}(2017)Fasshauer, F\"{o}rstel, Mucke, Arion,
  and Hergenhahn]{etmdexpt}
Fasshauer,~E.; F\"{o}rstel,~M.; Mucke,~M.; Arion,~T.; Hergenhahn,~U.
  Theoretical and Experimental Investigation of Electron Transfer Mediated
  Decay in ArKr Clusters. \emph{Chem. Phys.} \textbf{2017}, \emph{482},
  226--238\relax
\mciteBstWouldAddEndPuncttrue
\mciteSetBstMidEndSepPunct{\mcitedefaultmidpunct}
{\mcitedefaultendpunct}{\mcitedefaultseppunct}\relax
\EndOfBibitem
\bibitem[F\"{o}rstel \latin{et~al.}(2011)F\"{o}rstel, Mucke, Arion, Bradshaw,
  and Hergenhahn]{etmd2}
F\"{o}rstel,~M.; Mucke,~M.; Arion,~T.; Bradshaw,~A.~M.; Hergenhahn,~U.
  Autoionization Mediated by Electron Transfer. \emph{Phys. Rev. Lett.}
  \textbf{2011}, \emph{106}, 033402\relax
\mciteBstWouldAddEndPuncttrue
\mciteSetBstMidEndSepPunct{\mcitedefaultmidpunct}
{\mcitedefaultendpunct}{\mcitedefaultseppunct}\relax
\EndOfBibitem
\bibitem[Yan \latin{et~al.}(2013)Yan, Zhang, Ma, Xu, Li, Zhu, Feng, Zhang,
  Zhao, Zhang, Guo, and Liu]{etmdraregas}
Yan,~S.; Zhang,~P.; Ma,~X.; Xu,~S.; Li,~B.; Zhu,~X.~L.; Feng,~W.~T.;
  Zhang,~S.~F.; Zhao,~D.~M.; Zhang,~R.; Guo,~D.; Liu,~H.~P. Observation of
  Interatomic Coulombic Decay and Electron Transfer Mediated Decay in High
  Energy Electron Impact Ionization of Ar$_2$. \emph{Phys. Rev. A}
  \textbf{2013}, \emph{88}, 042712\relax
\mciteBstWouldAddEndPuncttrue
\mciteSetBstMidEndSepPunct{\mcitedefaultmidpunct}
{\mcitedefaultendpunct}{\mcitedefaultseppunct}\relax
\EndOfBibitem
\bibitem[Sakai \latin{et~al.}(2011)Sakai, Stoychev, Ouchi, Higuchi,
  Sch\"{o}ffler, Mazza, Fukuzawa, Nagaya, Yao, Tamenori, Kuleff, Saito, and
  Ueda]{etmdr1}
Sakai,~K.; Stoychev,~S.; Ouchi,~T.; Higuchi,~I.; Sch\"{o}ffler,~M.; Mazza,~T.;
  Fukuzawa,~H.; Nagaya,~K.; Yao,~M.; Tamenori,~Y.; Kuleff,~A.~I.; Saito,~N.;
  Ueda,~K. Electron Transfer Mediated Decay and Interatomic Coulombic Decay
  from the Triply Ionized States in Argon Dimers. \emph{Phys. Rev. Lett.}
  \textbf{2011}, \emph{106}, 033401\relax
\mciteBstWouldAddEndPuncttrue
\mciteSetBstMidEndSepPunct{\mcitedefaultmidpunct}
{\mcitedefaultendpunct}{\mcitedefaultseppunct}\relax
\EndOfBibitem
\bibitem[LaForge \latin{et~al.}(2016)LaForge, Stumpf, Gokhberg, von Vangerow,
  Stienkemeier, Kryzhevoi, O'Keeffe, Ciavardini, Krishnan, Coreno, Prince,
  Richter, Moshammer, Pfeifer, Cederbaum, and Mudrich]{hedroplt}
LaForge,~A.~C.; Stumpf,~V.; Gokhberg,~K.; von Vangerow,~J.; Stienkemeier,~F.;
  Kryzhevoi,~N.~V.; O'Keeffe,~P.; Ciavardini,~A.; Krishnan,~S.~R.; Coreno,~M.;
  Prince,~K.~C.; Richter,~R.; Moshammer,~R.; Pfeifer,~T.; Cederbaum,~L.~S.;
  Mudrich,~M. Enhanced Ionization of Embedded Clusters by
  Electron-Transfer-Mediated Decay in Helium Nanodroplets. \emph{Phys. Rev.
  Lett.} \textbf{2016}, \emph{116}, 203001\relax
\mciteBstWouldAddEndPuncttrue
\mciteSetBstMidEndSepPunct{\mcitedefaultmidpunct}
{\mcitedefaultendpunct}{\mcitedefaultseppunct}\relax
\EndOfBibitem
\bibitem[Ghosh \latin{et~al.}(2019)Ghosh, Cederbaum, and Gokhberg]{heli2}
Ghosh,~A.; Cederbaum,~L.~S.; Gokhberg,~K. Electron Transfer Mediated Decay in
  HeLi$_2$ Cluster: Potential Energy Surfaces and Decay Widths. \emph{J. Chem.
  Phys.} \textbf{2019}, \emph{150}, 164309\relax
\mciteBstWouldAddEndPuncttrue
\mciteSetBstMidEndSepPunct{\mcitedefaultmidpunct}
{\mcitedefaultendpunct}{\mcitedefaultseppunct}\relax
\EndOfBibitem
\bibitem[Kryzhevoi \latin{et~al.}(2007)Kryzhevoi, Averbukh, and
  Cederbaum]{hedrop}
Kryzhevoi,~N.~V.; Averbukh,~V.; Cederbaum,~L.~S. High Activity of Helium
  Droplets Following Ionization of Systems Inside Those Droplets. \emph{Phys.
  Rev. B} \textbf{2007}, \emph{76}, 094513\relax
\mciteBstWouldAddEndPuncttrue
\mciteSetBstMidEndSepPunct{\mcitedefaultmidpunct}
{\mcitedefaultendpunct}{\mcitedefaultseppunct}\relax
\EndOfBibitem
\bibitem[Ben~Ltaief \latin{et~al.}(2020)Ben~Ltaief, Shcherbinin, Mandal,
  Krishnan, Richter, Pfeifer, Bauer, Ghosh, Mudrich, Gokhberg, and
  LaForge]{etmdli}
Ben~Ltaief,~L.; Shcherbinin,~M.; Mandal,~S.; Krishnan,~S.~R.; Richter,~R.;
  Pfeifer,~T.; Bauer,~M.; Ghosh,~A.; Mudrich,~M.; Gokhberg,~K.; LaForge,~A.~C.
  Electron Transfer Mediated Decay of Alkali Dimers Attached to He
  Nanodroplets. \emph{Phys. Chem. Chem. Phys.} \textbf{2020}, \emph{22},
  8557--8564\relax
\mciteBstWouldAddEndPuncttrue
\mciteSetBstMidEndSepPunct{\mcitedefaultmidpunct}
{\mcitedefaultendpunct}{\mcitedefaultseppunct}\relax
\EndOfBibitem
\bibitem[Unger \latin{et~al.}(2017)Unger, Seidel, Th\"{u}rmer, Pohl, Aziz,
  Cederbaum, Muchov\'{a}, Slav\'{i}\u{c}ek, Winter, and Kryzhevoi]{unger}
Unger,~I.; Seidel,~R.; Th\"{u}rmer,~S.; Pohl,~M.~N.; Aziz,~E.~F.;
  Cederbaum,~L.~S.; Muchov\'{a},~E.; Slav\'{i}\u{c}ek,~P.; Winter,~B.;
  Kryzhevoi,~N.~V. Observation of Electron Transfer Mediated Decay in Aqueous
  Solution. \emph{Nat. Chem.} \textbf{2017}, \emph{9}, 708--714\relax
\mciteBstWouldAddEndPuncttrue
\mciteSetBstMidEndSepPunct{\mcitedefaultmidpunct}
{\mcitedefaultendpunct}{\mcitedefaultseppunct}\relax
\EndOfBibitem
\bibitem[M\"{u}ller and Cederbaum(2005)M\"{u}ller, and Cederbaum]{etmdhbond}
M\"{u}ller,~I.~B.; Cederbaum,~L.~S. Electronic Decay Following Ionization of
  Aqueous Li$^+$ Microsolvation Clusters. \emph{J. Chem. Phys.} \textbf{2005},
  \emph{122}, 094305\relax
\mciteBstWouldAddEndPuncttrue
\mciteSetBstMidEndSepPunct{\mcitedefaultmidpunct}
{\mcitedefaultendpunct}{\mcitedefaultseppunct}\relax
\EndOfBibitem
\bibitem[Stumpf \latin{et~al.}(2016)Stumpf, Gokhberg, and
  Cederbaum]{etmdhbond1}
Stumpf,~V.; Gokhberg,~K.; Cederbaum,~L.~S. The Role of Metal Ions in X-ray
  Induced Photochemistry. \emph{Nat. Chem.} \textbf{2016}, \emph{8},
  237--241\relax
\mciteBstWouldAddEndPuncttrue
\mciteSetBstMidEndSepPunct{\mcitedefaultmidpunct}
{\mcitedefaultendpunct}{\mcitedefaultseppunct}\relax
\EndOfBibitem
\bibitem[Ghosh \latin{et~al.}(2021)Ghosh, Cederbaum, and Gokhberg]{ghosh21}
Ghosh,~A.; Cederbaum,~L.~S.; Gokhberg,~K. Signature of the Neighbor's Quantum
  Nuclear Dynamics in the Electron Transfer Mediated Decay Spectra. \emph{Chem.
  Sci.} \textbf{2021}, \emph{12}, 9379--9385\relax
\mciteBstWouldAddEndPuncttrue
\mciteSetBstMidEndSepPunct{\mcitedefaultmidpunct}
{\mcitedefaultendpunct}{\mcitedefaultseppunct}\relax
\EndOfBibitem
\bibitem[Stumpf \latin{et~al.}(2016)Stumpf, Brunken, and Gokhberg]{microsolv}
Stumpf,~V.; Brunken,~C.; Gokhberg,~K. Impact of Metal Ion's Charge on the
  Interatomic Coulombic Decay Widths in Microsolvated Clusters. \emph{J. Chem.
  Phys.} \textbf{2016}, \emph{145}, 104306\relax
\mciteBstWouldAddEndPuncttrue
\mciteSetBstMidEndSepPunct{\mcitedefaultmidpunct}
{\mcitedefaultendpunct}{\mcitedefaultseppunct}\relax
\EndOfBibitem
\bibitem[Brun \latin{et~al.}(2009)Brun, Cloutier, Sicard-Roselli, Fromm, and
  Sanche]{brun}
Brun,~E.; Cloutier,~P.; Sicard-Roselli,~C.; Fromm,~M.; Sanche,~L. Damage
  Induced to DNA by Low-Energy (0-30 eV) Electrons under Vacuum and Atmospheric
  Conditions. \emph{J. Phys. Chem. B} \textbf{2009}, \emph{113},
  10008--10013\relax
\mciteBstWouldAddEndPuncttrue
\mciteSetBstMidEndSepPunct{\mcitedefaultmidpunct}
{\mcitedefaultendpunct}{\mcitedefaultseppunct}\relax
\EndOfBibitem
\bibitem[Alizadeh \latin{et~al.}(2015)Alizadeh, Orlando, and Sanche]{icdbiomol}
Alizadeh,~E.; Orlando,~T.~M.; Sanche,~L. Biomolecular Damage Induced by
  Ionizing Radiation: The Direct and Indirect Effects of Low-Energy Electrons
  on DNA. \emph{Annu. Rev. Phys. Chem.} \textbf{2015}, \emph{66},
  379--398\relax
\mciteBstWouldAddEndPuncttrue
\mciteSetBstMidEndSepPunct{\mcitedefaultmidpunct}
{\mcitedefaultendpunct}{\mcitedefaultseppunct}\relax
\EndOfBibitem
\bibitem[Ren \latin{et~al.}(2018)Ren, Wang, Skitnevskaya, Trofimov, Gokhberg,
  and Dorn]{icdbio2}
Ren,~X.; Wang,~E.; Skitnevskaya,~A.~D.; Trofimov,~A.~B.; Gokhberg,~K.; Dorn,~A.
  Experimental Evidence for Ultrafast Intermolecular Relaxation Processes in
  Hydrated Biomolecules. \emph{Nat. Phys.} \textbf{2018}, \emph{14},
  1062--1066\relax
\mciteBstWouldAddEndPuncttrue
\mciteSetBstMidEndSepPunct{\mcitedefaultmidpunct}
{\mcitedefaultendpunct}{\mcitedefaultseppunct}\relax
\EndOfBibitem
\bibitem[Sussman and Weinstein(1989)Sussman, and Weinstein]{enzyme}
Sussman,~F.; Weinstein,~H. On the Ion Selectivity in Ca-Binding Proteins: The
  Cyclo(-L-Pro-Gly-)3 Peptide as a Model. \emph{Proc. Natl. Acad. Sci. USA}
  \textbf{1989}, \emph{86}, 7880--7884\relax
\mciteBstWouldAddEndPuncttrue
\mciteSetBstMidEndSepPunct{\mcitedefaultmidpunct}
{\mcitedefaultendpunct}{\mcitedefaultseppunct}\relax
\EndOfBibitem
\bibitem[Dill(1990)]{proteinfolding}
Dill,~K.~A. Dominant Forces in Protein Folding. \emph{Biochemistry}
  \textbf{1990}, \emph{29}, 7133--7155\relax
\mciteBstWouldAddEndPuncttrue
\mciteSetBstMidEndSepPunct{\mcitedefaultmidpunct}
{\mcitedefaultendpunct}{\mcitedefaultseppunct}\relax
\EndOfBibitem
\bibitem[Forrest(2014)]{brainsignals}
Forrest,~M.~D. The Sodium-Potassium Pump is an Information Processing Element
  in Brain Computation. \emph{Front. Physiol.} \textbf{2014}, \emph{5},
  472\relax
\mciteBstWouldAddEndPuncttrue
\mciteSetBstMidEndSepPunct{\mcitedefaultmidpunct}
{\mcitedefaultendpunct}{\mcitedefaultseppunct}\relax
\EndOfBibitem
\bibitem[Clausen \latin{et~al.}(2017)Clausen, Hilbers, and Poulsen]{braindis}
Clausen,~M.~V.; Hilbers,~F.; Poulsen,~H. The Structure and Function of the
  Na,K-ATPase Isoforms in Health and Disease. \emph{Front. Physiol.}
  \textbf{2017}, \emph{8}, 371\relax
\mciteBstWouldAddEndPuncttrue
\mciteSetBstMidEndSepPunct{\mcitedefaultmidpunct}
{\mcitedefaultendpunct}{\mcitedefaultseppunct}\relax
\EndOfBibitem
\bibitem[Ghosh and Vaval(2014)Ghosh, and Vaval]{capeom}
Ghosh,~A.; Vaval,~N. Geometry-Dependent Lifetime of Interatomic Coulombic Decay
  using Equation-of-Motion Coupled Cluster Method. \emph{J. Chem. Phys.}
  \textbf{2014}, \emph{141}, 234108\relax
\mciteBstWouldAddEndPuncttrue
\mciteSetBstMidEndSepPunct{\mcitedefaultmidpunct}
{\mcitedefaultendpunct}{\mcitedefaultseppunct}\relax
\EndOfBibitem
\bibitem[Ghosh \latin{et~al.}(2013)Ghosh, Pal, and Vaval]{capeom1}
Ghosh,~A.; Pal,~S.; Vaval,~N. Study of Interatomic Coulombic Decay of
  Ne(H2O)$_n$ (n=1,3) Clusters Using Equation-of-Motion Coupled Cluster Method.
  \emph{J. Chem. Phys.} \textbf{2013}, \emph{139}, 064112\relax
\mciteBstWouldAddEndPuncttrue
\mciteSetBstMidEndSepPunct{\mcitedefaultmidpunct}
{\mcitedefaultendpunct}{\mcitedefaultseppunct}\relax
\EndOfBibitem
\bibitem[Ghosh \latin{et~al.}(2014)Ghosh, Vaval, Pal, and Bartlett]{capeom2}
Ghosh,~A.; Vaval,~N.; Pal,~S.; Bartlett,~R.~J. Complex Absorbing Potential
  Based Equation-of-Motion Coupled Cluster Method for the Potential Energy
  Curve of CO$^-_2$ Anion. \emph{J. Chem. Phys.} \textbf{2014}, \emph{141},
  164113\relax
\mciteBstWouldAddEndPuncttrue
\mciteSetBstMidEndSepPunct{\mcitedefaultmidpunct}
{\mcitedefaultendpunct}{\mcitedefaultseppunct}\relax
\EndOfBibitem
\bibitem[Jagau \latin{et~al.}(2014)Jagau, Zuev, Bravaya, Epifanovsky, and
  Krylov]{capeom3a}
Jagau,~T.~C.; Zuev,~D.; Bravaya,~K.~B.; Epifanovsky,~E.; Krylov,~A.~I. A Fresh
  Look at Resonances and Complex Absorbing Potentials: Density Matrix-Based
  Approach. \emph{J. Phys. Chem. Lett.} \textbf{2014}, \emph{5}, 310--315\relax
\mciteBstWouldAddEndPuncttrue
\mciteSetBstMidEndSepPunct{\mcitedefaultmidpunct}
{\mcitedefaultendpunct}{\mcitedefaultseppunct}\relax
\EndOfBibitem
\bibitem[Zuev \latin{et~al.}(2014)Zuev, Jagau, Bravaya, Epifanovsky, Shao,
  Sundstrom, Head-Gordon, and Krylov]{capeom3b}
Zuev,~D.; Jagau,~T.~C.; Bravaya,~K.~B.; Epifanovsky,~E.; Shao,~Y.;
  Sundstrom,~E.; Head-Gordon,~M.; Krylov,~A.~I. Complex Absorbing Potentials
  within EOM-CC Family of Methods: Theory, Implementation, and Benchmarks.
  \emph{J. Chem. Phys.} \textbf{2014}, \emph{141}, 024102\relax
\mciteBstWouldAddEndPuncttrue
\mciteSetBstMidEndSepPunct{\mcitedefaultmidpunct}
{\mcitedefaultendpunct}{\mcitedefaultseppunct}\relax
\EndOfBibitem
\bibitem[Jagau and Krylov(2014)Jagau, and Krylov]{capeom3c}
Jagau,~T.~C.; Krylov,~A.~I. Complex Absorbing Potential Equation-of-Motion
  Coupled-Cluster Method Yields Smooth and Internally Consistent Potential
  Energy Surfaces and Lifetimes for Molecular Resonances. \emph{J. Phys. Chem.
  Lett.} \textbf{2014}, \emph{5}, 3078--3085\relax
\mciteBstWouldAddEndPuncttrue
\mciteSetBstMidEndSepPunct{\mcitedefaultmidpunct}
{\mcitedefaultendpunct}{\mcitedefaultseppunct}\relax
\EndOfBibitem
\bibitem[Sajeev \latin{et~al.}(2014)Sajeev, Ghosh, Vaval, and Pal]{capeom4}
Sajeev,~Y.; Ghosh,~A.; Vaval,~N.; Pal,~S. Coupled Cluster Methods for
  Autoionisation Resonances. \emph{Int. Rev. Phys. Chem.} \textbf{2014},
  \emph{33}, 397--425\relax
\mciteBstWouldAddEndPuncttrue
\mciteSetBstMidEndSepPunct{\mcitedefaultmidpunct}
{\mcitedefaultendpunct}{\mcitedefaultseppunct}\relax
\EndOfBibitem
\bibitem[Bartlett and Musia\l(2007)Bartlett, and Musia\l]{eomcc}
Bartlett,~R.~J.; Musia\l,~M. Coupled Cluster Theory in Quantum Chemistry.
  \emph{Rev. Mod. Phys.} \textbf{2007}, \emph{79}, 291--352\relax
\mciteBstWouldAddEndPuncttrue
\mciteSetBstMidEndSepPunct{\mcitedefaultmidpunct}
{\mcitedefaultendpunct}{\mcitedefaultseppunct}\relax
\EndOfBibitem
\bibitem[Lyakh \latin{et~al.}(2012)Lyakh, Musia\l, Lotrich, and
  Bartlett]{eomcc1}
Lyakh,~D.~I.; Musia\l,~M.; Lotrich,~V.~F.; Bartlett,~R.~J. Multireference
  Nature of Chemistry: The Coupled Cluster View. \emph{Chem. Rev.}
  \textbf{2012}, \emph{112}, 182--243\relax
\mciteBstWouldAddEndPuncttrue
\mciteSetBstMidEndSepPunct{\mcitedefaultmidpunct}
{\mcitedefaultendpunct}{\mcitedefaultseppunct}\relax
\EndOfBibitem
\bibitem[Nooijen and Bartlett(1995)Nooijen, and Bartlett]{eomcc2}
Nooijen,~M.; Bartlett,~R.~J. Equation-of-Motion Coupled Cluster Method for
  Electron Attachment. \emph{J. Chem. Phys.} \textbf{1995}, \emph{102},
  3629--3647\relax
\mciteBstWouldAddEndPuncttrue
\mciteSetBstMidEndSepPunct{\mcitedefaultmidpunct}
{\mcitedefaultendpunct}{\mcitedefaultseppunct}\relax
\EndOfBibitem
\bibitem[Stanton and Bartlett(1993)Stanton, and Bartlett]{eomcc3}
Stanton,~J.~F.; Bartlett,~R.~J. The Equation-of-Motion Coupled Cluster Method.
  A Systematic Biorthogonal Approach to Molecular Excitation Energies,
  Transition Probabilities, and Excited State Properties. \emph{J. Chem. Phys.}
  \textbf{1993}, \emph{98}, 7029--7039\relax
\mciteBstWouldAddEndPuncttrue
\mciteSetBstMidEndSepPunct{\mcitedefaultmidpunct}
{\mcitedefaultendpunct}{\mcitedefaultseppunct}\relax
\EndOfBibitem
\bibitem[Nooijen and Bartlett(1997)Nooijen, and Bartlett]{eomcc4}
Nooijen,~M.; Bartlett,~R.~J. Similarity Transformed Equation-of-Motion
  Coupled-Cluster Theory: Details, Examples, and Comparisons. \emph{J. Chem.
  Phys.} \textbf{1997}, \emph{107}, 6812--6830\relax
\mciteBstWouldAddEndPuncttrue
\mciteSetBstMidEndSepPunct{\mcitedefaultmidpunct}
{\mcitedefaultendpunct}{\mcitedefaultseppunct}\relax
\EndOfBibitem
\bibitem[Musia\l and Bartlett(2011)Musia\l, and Bartlett]{eomcc5}
Musia\l,~M.; Bartlett,~R.~J. Charge Transfer Separability and Size Extensivity
  in the Equation-of-Motion Coupled Cluster Method: EOM-CCx. \emph{J. Chem.
  Phys.} \textbf{2011}, \emph{134}, 034106\relax
\mciteBstWouldAddEndPuncttrue
\mciteSetBstMidEndSepPunct{\mcitedefaultmidpunct}
{\mcitedefaultendpunct}{\mcitedefaultseppunct}\relax
\EndOfBibitem
\bibitem[Musia\l \latin{et~al.}(2003)Musia\l, Kucharski, and Bartlett]{eomcc6}
Musia\l,~M.; Kucharski,~S.~A.; Bartlett,~R.~J. Equation-of-Motion Coupled
  Cluster Method with Full Inclusion of the Connected Triple Excitations for
  Ionized States: IP-EOM-CCSDT. \emph{J. Chem. Phys.} \textbf{2003},
  \emph{118}, 1128--1136\relax
\mciteBstWouldAddEndPuncttrue
\mciteSetBstMidEndSepPunct{\mcitedefaultmidpunct}
{\mcitedefaultendpunct}{\mcitedefaultseppunct}\relax
\EndOfBibitem
\bibitem[Siegert(1939)]{siegert}
Siegert,~A. J.~F. On the Derivation of the Dispersion Formula for Nuclear
  Reactions. \emph{Phys. Rev.} \textbf{1939}, \emph{56}, 750--752\relax
\mciteBstWouldAddEndPuncttrue
\mciteSetBstMidEndSepPunct{\mcitedefaultmidpunct}
{\mcitedefaultendpunct}{\mcitedefaultseppunct}\relax
\EndOfBibitem
\bibitem[Moiseyev and Corcoran(1979)Moiseyev, and Corcoran]{comscal}
Moiseyev,~N.; Corcoran,~C. Autoionizing States of ${\mathrm{H}}_{2}$ and
  ${\mathrm{H}}_{2}^{\ensuremath{-}}$ using the Complex-Scaling Method.
  \emph{Phys. Rev. A} \textbf{1979}, \emph{20}, 814--817\relax
\mciteBstWouldAddEndPuncttrue
\mciteSetBstMidEndSepPunct{\mcitedefaultmidpunct}
{\mcitedefaultendpunct}{\mcitedefaultseppunct}\relax
\EndOfBibitem
\bibitem[Moiseyev(1998)]{comscal1}
Moiseyev,~N. Quantum Theory of Resonances: Calculating Energies, Widths and
  Cross-Sections by Complex Scaling. \emph{Phys. Rep.} \textbf{1998},
  \emph{302}, 212--293\relax
\mciteBstWouldAddEndPuncttrue
\mciteSetBstMidEndSepPunct{\mcitedefaultmidpunct}
{\mcitedefaultendpunct}{\mcitedefaultseppunct}\relax
\EndOfBibitem
\bibitem[Moiseyev(1998)]{comscal2}
Moiseyev,~N. Derivations of Universal Exact Complex Absorption Potentials by
  the Generalized Complex Coordinate Method. \emph{J. Phys. B: At. Mol. Opt.
  Phys.} \textbf{1998}, \emph{31}, 1431--1441\relax
\mciteBstWouldAddEndPuncttrue
\mciteSetBstMidEndSepPunct{\mcitedefaultmidpunct}
{\mcitedefaultendpunct}{\mcitedefaultseppunct}\relax
\EndOfBibitem
\bibitem[Sommerfeld \latin{et~al.}(1998)Sommerfeld, Riss, Meyer, Cederbaum,
  Engels, and Suter]{cap1}
Sommerfeld,~T.; Riss,~U.~V.; Meyer,~H.~D.; Cederbaum,~L.~S.; Engels,~B.;
  Suter,~H.~U. Temporary Anions - Calculation of Energy and Lifetime by
  Absorbing Potentials: The Resonance. \emph{J. Phys. B: At. Mol. Opt. Phys.}
  \textbf{1998}, \emph{31}, 4107--4122\relax
\mciteBstWouldAddEndPuncttrue
\mciteSetBstMidEndSepPunct{\mcitedefaultmidpunct}
{\mcitedefaultendpunct}{\mcitedefaultseppunct}\relax
\EndOfBibitem
\bibitem[Jolicard and Austin(1985)Jolicard, and Austin]{cap2}
Jolicard,~G.; Austin,~E.~J. Optical Potential Stabilisation Method for
  Predicting Resonance Levels. \emph{Chem. Phys. Lett.} \textbf{1985},
  \emph{121}, 106--110\relax
\mciteBstWouldAddEndPuncttrue
\mciteSetBstMidEndSepPunct{\mcitedefaultmidpunct}
{\mcitedefaultendpunct}{\mcitedefaultseppunct}\relax
\EndOfBibitem
\bibitem[Riss and Meyer(1993)Riss, and Meyer]{cap3}
Riss,~U.~V.; Meyer,~H.~D. Calculation of Resonance Energies and Widths using
  the Complex Absorbing Potential Method. \emph{J. Phys. B: At. Mol. Opt.
  Phys.} \textbf{1993}, \emph{26}, 4503--4535\relax
\mciteBstWouldAddEndPuncttrue
\mciteSetBstMidEndSepPunct{\mcitedefaultmidpunct}
{\mcitedefaultendpunct}{\mcitedefaultseppunct}\relax
\EndOfBibitem
\bibitem[Muga \latin{et~al.}(2004)Muga, Palao, Navarro, and Egusquiza]{cap4}
Muga,~J.; Palao,~J.; Navarro,~B.; Egusquiza,~I. Complex Absorbing Potentials.
  \emph{Phys. Rep.} \textbf{2004}, \emph{395}, 357--426\relax
\mciteBstWouldAddEndPuncttrue
\mciteSetBstMidEndSepPunct{\mcitedefaultmidpunct}
{\mcitedefaultendpunct}{\mcitedefaultseppunct}\relax
\EndOfBibitem
\bibitem[Riss and Meyer(1998)Riss, and Meyer]{cap6}
Riss,~U.~V.; Meyer,~H.~D. The Transformative Complex Absorbing Potential
  Method: A Bridge between Complex Absorbing Potentials and Smooth Exterior
  Scaling. \emph{J. Phys. B: At. Mol. Opt. Phys.} \textbf{1998}, \emph{31},
  2279--2304\relax
\mciteBstWouldAddEndPuncttrue
\mciteSetBstMidEndSepPunct{\mcitedefaultmidpunct}
{\mcitedefaultendpunct}{\mcitedefaultseppunct}\relax
\EndOfBibitem
\bibitem[Sajeev \latin{et~al.}(2005)Sajeev, Santra, and Pal]{capfsa}
Sajeev,~Y.; Santra,~R.; Pal,~S. Correlated Complex Independent Particle
  Potential for Calculating Electronic Resonances. \emph{J. Chem. Phys.}
  \textbf{2005}, \emph{123}, 204110\relax
\mciteBstWouldAddEndPuncttrue
\mciteSetBstMidEndSepPunct{\mcitedefaultmidpunct}
{\mcitedefaultendpunct}{\mcitedefaultseppunct}\relax
\EndOfBibitem
\bibitem[Sajeev and Pal(2005)Sajeev, and Pal]{capfsb}
Sajeev,~Y.; Pal,~S. A General Formalism of the Fock Space Multireference
  Coupled Cluster Method for Investigating Molecular Electronic Resonances.
  \emph{Mol. Phys.} \textbf{2005}, \emph{103}, 2267--2275\relax
\mciteBstWouldAddEndPuncttrue
\mciteSetBstMidEndSepPunct{\mcitedefaultmidpunct}
{\mcitedefaultendpunct}{\mcitedefaultseppunct}\relax
\EndOfBibitem
\bibitem[Ghosh \latin{et~al.}(2013)Ghosh, Karne, Pal, and Vaval]{pccparyya}
Ghosh,~A.; Karne,~A.; Pal,~S.; Vaval,~N. CAP/EOM-CCSD Method for the Study of
  Potential Curves of Resonant States. \emph{Phys. Chem. Chem. Phys.}
  \textbf{2013}, \emph{15}, 17915--17921\relax
\mciteBstWouldAddEndPuncttrue
\mciteSetBstMidEndSepPunct{\mcitedefaultmidpunct}
{\mcitedefaultendpunct}{\mcitedefaultseppunct}\relax
\EndOfBibitem
\bibitem[Tomasi \latin{et~al.}(2005)Tomasi, Mennucci, and Cammi]{pcm}
Tomasi,~J.; Mennucci,~B.; Cammi,~R. Quantum Mechanical Continuum Solvation
  Models. \emph{Chem. Rev.} \textbf{2005}, \emph{105}, 2999--3094\relax
\mciteBstWouldAddEndPuncttrue
\mciteSetBstMidEndSepPunct{\mcitedefaultmidpunct}
{\mcitedefaultendpunct}{\mcitedefaultseppunct}\relax
\EndOfBibitem
\bibitem[Frisch \latin{et~al.}(2009)Frisch, Trucks, Schlegel, Scuseria, Robb,
  Cheeseman, Scalmani, Barone, Mennucci, Petersson, Nakatsuji, Caricato, Li,
  Hratchian, Izmaylov, Bloino, Zheng, Sonnenberg, Hada, Ehara, Toyota, Fukuda,
  Hasegawa, Ishida, Nakajima, Honda, Kitao, Nakai, Vreven, Montgomery, Peralta,
  Ogliaro, Bearpark, Heyd, Brothers, Kudin, Staroverov, Kobayashi, Norm~and,
  Raghavachari, Rendell, Burant, Iyengar, Tomasi, Cossi, Rega, Millam, Klene,
  Knox, Cross, Bakken, Adamo, Jaramillo, Gomperts, Stratmann, Yazyev, Austin,
  Cammi, Pomelli, Ochterski, Martin, Morokuma, Zakrzewski, Voth, Salvador,
  Dannenberg, Dapprich, Daniels, Farkas, Foresman, Ortiz, Cioslowski, and
  Fox]{g09}
Frisch,~M.~J.; Trucks,~G.~W.; Schlegel,~H.~B.; Scuseria,~G.~E.; Robb,~M.~A.;
  Cheeseman,~J.~R.; Scalmani,~G.; Barone,~V.; Mennucci,~B.; Petersson,~G.~A.;
  Nakatsuji,~H.; Caricato,~M.; Li,~X.; Hratchian,~H.~P.; Izmaylov,~A.~F.;
  Bloino,~J.; Zheng,~G.; Sonnenberg,~J.~L.; Hada,~M.; Ehara,~M.; Toyota,~K.;
  Fukuda,~R.; Hasegawa,~J.; Ishida,~M.; Nakajima,~T.; Honda,~Y.; Kitao,~O.;
  Nakai,~H.; Vreven,~T.; Montgomery,~J.,~J.~A.; Peralta,~J.~E.; Ogliaro,~F.;
  Bearpark,~M.; Heyd,~J.~J.; Brothers,~E.; Kudin,~K.~N.; Staroverov,~V.~N.;
  Kobayashi,~R.; Norm~and,~J.; Raghavachari,~K.; Rendell,~A.; Burant,~J.~C.;
  Iyengar,~S.~S.; Tomasi,~J.; Cossi,~M.; Rega,~N.; Millam,~J.~M.; Klene,~M.;
  Knox,~J.~E.; Cross,~J.~B.; Bakken,~V.; Adamo,~C.; Jaramillo,~J.;
  Gomperts,~R.; Stratmann,~R.~E.; Yazyev,~O.; Austin,~A.~J.; Cammi,~R.;
  Pomelli,~C.; Ochterski,~J.~W.; Martin,~R.~L.; Morokuma,~K.;
  Zakrzewski,~V.~G.; Voth,~G.~A.; Salvador,~P.; Dannenberg,~J.~J.;
  Dapprich,~S.; Daniels,~A.~D.; Farkas,~Ã.; Foresman,~J.~B.; Ortiz,~J.~V.;
  Cioslowski,~J.; Fox,~D.~J. \emph{Gaussian-09}, {R}evision {C}.01 ed.;
  Gaussian Inc.: Wallingford CT, 2009\relax
\mciteBstWouldAddEndPuncttrue
\mciteSetBstMidEndSepPunct{\mcitedefaultmidpunct}
{\mcitedefaultendpunct}{\mcitedefaultseppunct}\relax
\EndOfBibitem
\bibitem[Dunning(1989)]{ccpvdz}
Dunning,~T.~H. Gaussian Basis Sets for Use in Correlated Molecular
  Calculations. I. The Atoms Boron Through Neon and Hydrogen. \emph{J. Chem.
  Phys.} \textbf{1989}, \emph{90}, 1007--1023\relax
\mciteBstWouldAddEndPuncttrue
\mciteSetBstMidEndSepPunct{\mcitedefaultmidpunct}
{\mcitedefaultendpunct}{\mcitedefaultseppunct}\relax
\EndOfBibitem
\bibitem[Kaufmann \latin{et~al.}(1989)Kaufmann, Baumeister, and
  Jungen]{Kaufmann}
Kaufmann,~K.; Baumeister,~W.; Jungen,~M. Universal Gaussian Basis Sets for an
  Optimum Representation of Rydberg and Continuum Wavefunctions. \emph{J. Phys.
  B: At. Mol. Opt. Phys.} \textbf{1989}, \emph{22}, 2223--2240\relax
\mciteBstWouldAddEndPuncttrue
\mciteSetBstMidEndSepPunct{\mcitedefaultmidpunct}
{\mcitedefaultendpunct}{\mcitedefaultseppunct}\relax
\EndOfBibitem
\bibitem[Kendall \latin{et~al.}(1992)Kendall, Dunning, and Harrison]{ccpvtz}
Kendall,~R.~A.; Dunning,~T.~H.; Harrison,~R.~J. Electron Affinities of the
  First-Row Atoms Revisited. Systematic Basis Sets and Wave Functions. \emph{J.
  Chem. Phys.} \textbf{1992}, \emph{96}, 6796--6806\relax
\mciteBstWouldAddEndPuncttrue
\mciteSetBstMidEndSepPunct{\mcitedefaultmidpunct}
{\mcitedefaultendpunct}{\mcitedefaultseppunct}\relax
\EndOfBibitem
\bibitem[Becke(1993)]{B3}
Becke,~A.~D. Density-Functional Thermochemistry. III. The Role of Exact
  Exchange. \emph{J. Chem. Phys.} \textbf{1993}, \emph{98}, 5648--5652\relax
\mciteBstWouldAddEndPuncttrue
\mciteSetBstMidEndSepPunct{\mcitedefaultmidpunct}
{\mcitedefaultendpunct}{\mcitedefaultseppunct}\relax
\EndOfBibitem
\bibitem[Lee \latin{et~al.}(1988)Lee, Yang, and Parr]{LYP}
Lee,~C.; Yang,~W.; Parr,~R.~G. Development of the Colle-Salvetti
  Correlation-Energy Formula into a Functional of the Electron Density.
  \emph{Phys. Rev. B} \textbf{1988}, \emph{37}, 785--789\relax
\mciteBstWouldAddEndPuncttrue
\mciteSetBstMidEndSepPunct{\mcitedefaultmidpunct}
{\mcitedefaultendpunct}{\mcitedefaultseppunct}\relax
\EndOfBibitem
\bibitem[Stephens \latin{et~al.}(1994)Stephens, Devlin, Chabalowski, and
  Frisch]{b3lyp}
Stephens,~P.~J.; Devlin,~F.~J.; Chabalowski,~C.~F.; Frisch,~M.~J. Ab Initio
  Calculation of Vibrational Absorption and Circular Dichroism Spectra Using
  Density Functional Force Fields. \emph{J. Phys. Chem.} \textbf{1994},
  \emph{98}, 11623--11627\relax
\mciteBstWouldAddEndPuncttrue
\mciteSetBstMidEndSepPunct{\mcitedefaultmidpunct}
{\mcitedefaultendpunct}{\mcitedefaultseppunct}\relax
\EndOfBibitem
\bibitem[Vosko \latin{et~al.}(1980)Vosko, Wilk, and Nusair]{VWN}
Vosko,~S.~H.; Wilk,~L.; Nusair,~M. Accurate Spin-Dependent Electron Liquid
  Correlation Energies for Local Spin Density Calculations: A Critical
  Analysis. \emph{Can. J. Phys.} \textbf{1980}, \emph{58}, 1200--1211\relax
\mciteBstWouldAddEndPuncttrue
\mciteSetBstMidEndSepPunct{\mcitedefaultmidpunct}
{\mcitedefaultendpunct}{\mcitedefaultseppunct}\relax
\EndOfBibitem
\bibitem[M\"{u}ller \latin{et~al.}(2004)M\"{u}ller, Cederbaum, and
  Tarantelli]{imke2}
M\"{u}ller,~I.~B.; Cederbaum,~L.~S.; Tarantelli,~F. Microsolvation of Li$^+$ in
  Water Analyzed by Ionization and Double Ionization. \emph{J. Phys. Chem. A}
  \textbf{2004}, \emph{108}, 5831--5844\relax
\mciteBstWouldAddEndPuncttrue
\mciteSetBstMidEndSepPunct{\mcitedefaultmidpunct}
{\mcitedefaultendpunct}{\mcitedefaultseppunct}\relax
\EndOfBibitem
\bibitem[Papajak \latin{et~al.}(2009)Papajak, Leverentz, Zheng, and
  Truhlar]{minmalaug}
Papajak,~E.; Leverentz,~H.~R.; Zheng,~J.; Truhlar,~D.~G. Efficient Diffuse
  Basis Sets: cc-pVxZ+ and maug-cc-pVxZ. \emph{J. Chem. Theory Comput.}
  \textbf{2009}, \emph{5}, 3330--3330\relax
\mciteBstWouldAddEndPuncttrue
\mciteSetBstMidEndSepPunct{\mcitedefaultmidpunct}
{\mcitedefaultendpunct}{\mcitedefaultseppunct}\relax
\EndOfBibitem
\bibitem[Pritchard \latin{et~al.}(2019)Pritchard, Altarawy, Didier, Gibson, and
  Windus]{esml}
Pritchard,~B.~P.; Altarawy,~D.; Didier,~B.; Gibson,~T.~D.; Windus,~T.~L. New
  Basis Set Exchange: An Open, Up-to-Date Resource for the Molecular Sciences
  Community. \emph{J. Chem. Inf. Model.} \textbf{2019}, \emph{59},
  4814--4820\relax
\mciteBstWouldAddEndPuncttrue
\mciteSetBstMidEndSepPunct{\mcitedefaultmidpunct}
{\mcitedefaultendpunct}{\mcitedefaultseppunct}\relax
\EndOfBibitem
\bibitem[\"{O}hrwall \latin{et~al.}(2010)\"{O}hrwall, Ottosson, Pokapanich,
  Legendre, Svensson, and Bj\"{o}rneholm]{Nawaterexpt}
\"{O}hrwall,~G.; Ottosson,~N.; Pokapanich,~W.; Legendre,~S.; Svensson,~S.;
  Bj\"{o}rneholm,~O. Charge Dependence of Solvent-Mediated Intermolecular
  Coster-Kronig Decay Dynamics of Aqueous Ions. \emph{J. Phys. Chem. B}
  \textbf{2010}, \emph{114}, 17057--17061\relax
\mciteBstWouldAddEndPuncttrue
\mciteSetBstMidEndSepPunct{\mcitedefaultmidpunct}
{\mcitedefaultendpunct}{\mcitedefaultseppunct}\relax
\EndOfBibitem
\bibitem[McLean and Chandler(1980)McLean, and Chandler]{Na2Ar6311g}
McLean,~A.~D.; Chandler,~G.~S. Contracted Gaussian Basis Sets for Molecular
  Calculations. I. Second Row Atoms, Z=11-18. \emph{J. Chem. Phys.}
  \textbf{1980}, \emph{72}, 5639--5648\relax
\mciteBstWouldAddEndPuncttrue
\mciteSetBstMidEndSepPunct{\mcitedefaultmidpunct}
{\mcitedefaultendpunct}{\mcitedefaultseppunct}\relax
\EndOfBibitem
\bibitem[Blaudeau \latin{et~al.}(1997)Blaudeau, McGrath, Curtiss, and
  Radom]{K6311g}
Blaudeau,~J.-P.; McGrath,~M.~P.; Curtiss,~L.~A.; Radom,~L. Extension of
  Gaussian-2 (G2) Theory to Molecules Containing Third-Row Atoms K and Ca.
  \emph{J. Chem. Phys.} \textbf{1997}, \emph{107}, 5016--5021\relax
\mciteBstWouldAddEndPuncttrue
\mciteSetBstMidEndSepPunct{\mcitedefaultmidpunct}
{\mcitedefaultendpunct}{\mcitedefaultseppunct}\relax
\EndOfBibitem
\bibitem[Hill and Peterson(2017)Hill, and Peterson]{x2cbasis}
Hill,~J.~G.; Peterson,~K.~A. Gaussian Basis Sets for Use in Correlated
  Molecular Calculations. XI. Pseudopotential-Based and All-Electron
  Relativistic Basis Sets for Alkali Metal (K-Fr) and Alkaline Earth (Ca-Ra)
  Elements. \emph{J. Chem. Phys.} \textbf{2017}, \emph{147}, 244106\relax
\mciteBstWouldAddEndPuncttrue
\mciteSetBstMidEndSepPunct{\mcitedefaultmidpunct}
{\mcitedefaultendpunct}{\mcitedefaultseppunct}\relax
\EndOfBibitem
\bibitem[Pokapanich \latin{et~al.}(2009)Pokapanich, Bergersen, Bradeanu,
  Marinho, Lindblad, Legendre, Rosso, Svensson, ~, Tchaplyguine, ~, Kryzhevoi,
  and Cederbaum]{kcl-water}
Pokapanich,~W.; Bergersen,~H.; Bradeanu,~I.~L.; Marinho,~R. R.~T.;
  Lindblad,~A.; Legendre,~S.; Rosso,~A.; Svensson,~S.; ~,~O.~B.;
  Tchaplyguine,~M.; ~,~G.~Ã.; Kryzhevoi,~N.~V.; Cederbaum,~L.~S. Auger Electron
  Spectroscopy as a Probe of the Solution of Aqueous Ions. \emph{J. Am. Chem.
  Soc.} \textbf{2009}, \emph{131}, 7264--7271\relax
\mciteBstWouldAddEndPuncttrue
\mciteSetBstMidEndSepPunct{\mcitedefaultmidpunct}
{\mcitedefaultendpunct}{\mcitedefaultseppunct}\relax
\EndOfBibitem
\end{mcitethebibliography}
\end{document}